\documentclass[twocolumn]{aastex631}
\usepackage{natbib}
\usepackage{appendix}
\usepackage{multirow}
\usepackage{longtable}
\newcommand{\tnt}{\textsc{Dynamite}}

\defcitealias{Dietrich2020}{Paper I}
\defcitealias{Dietrich2021}{Paper II}
\defcitealias{Dietrich2022}{Paper III}
\defcitealias{Basant2022a}{Paper IV}

\begin{document}

\title{Assessing Exoplanetary System Architectures with \textsc{DYNAMITE} Including Observational Upper Limits}

\correspondingauthor{Jamie Dietrich}
\email{jdietrich@asu.edu}

\author[0000-0001-6320-7410]{Jamie Dietrich}
\affiliation{School of Earth and Space Exploration, Arizona State University, Tempe, AZ 85287, USA}





\begin{abstract}
The information gathered from observing planetary systems is not limited to the discovery of planets, but also includes the observational upper limits constraining the presence of any additional planets. Incorporating these upper limits into statistical analyses of individual systems can significantly improve our ability to find hidden planets in these systems by narrowing the parameter space in which to search. Here I include radial velocity (RV), transit, and transit timing variation (TTV) upper limits on additional planets in known multi-planet systems into the \tnt{} software package and test their impact on the predicted planets for these systems. The tests are run on systems with previous \tnt{} analysis and with updated known planet parameters in the 2--3 years since the original predictions. I find that the RV limits provide the strongest constraints on additional planets, lowering the likelihood of finding them within orbital periods of $\sim$10--100 days in the inner planetary systems, as well as truncating the likely planet size (radius and/or mass) distributions towards planets smaller than those currently observed. Transit and TTV limits also provide information on the size and inclination distributions of both the known and predicted planets in the system. Utilizing these limits on a wider range of systems in the near future will help determine which systems might be able to host temperate terrestrial planets and contribute to the search for extraterrestrial biosignatures.
\end{abstract}



\section{Introduction} \label{sec:intro}

One of the main contributors to the search for extrasolar planets, and the first to produce robust and accurate results, was the radial velocity method. This was first proposed as far back as the 1950s by \citet[][]{Struve1952}, and in the late 1980s and early 1990s instruments such as ELODIE \citep[][]{Baranne1996} were able to have the precision required to definitively discover the first planetary-mass companion around a main-sequence star \citep[][]{Mayor1995}. As a major driver of both public and scientific interest, the focus on the search for additional planets outside the solar system produced great improvements in the technology needed for high-precision and high-accuracy observations. RV observations from spectrographs like HARPS \citep[][]{Mayor2003}, ESPRESSO \citep[][]{Pepe2010}, EXPRES \citep[][]{Jurgenson2016}, and NEID \citep[][]{Schwab2016} have routinely delivered sub-m/s precision, even getting down to 30 cm/s in some cases, which would be able to pick up planets only a few times larger than Earth in the habitable zone of Sun-like stars.

Since the discovery of the first planet to transit its host star \citep[][]{Charbonneau2000, Henry2000}, the transit method has become the most productive method of discovering planets, especially with the Kepler Space Telescope \citep[][]{Borucki2010} and its multi-year surveys in its prime and extended missions. Kepler was able to produce a statistically robust population of exoplanets with orbital periods less than two years around FGK stars, as well as provide detailed information on planetary system architectures and the trends therein. In addition, transit observations are precise enough to measure slight timing variations caused by the gravitational perturbations of an additional planet pulling on the transiting planet \citep[][]{Holman2005}. The Transiting Exoplanet Survey Satellite \citep[TESS;][]{Ricker2015} has the sensitivity to detect TTVs down to a few minutes for thousands of stars all across the sky \citep[e.g.,][]{Hadden2019}.

Planets can be detected and confirmed with multiple observation methods, but the information gathered from observing a planetary system is not limited to these observations. Non-detections of planets, combined with understanding the detection limits of the observation method, provide observational upper limits on the presence of additional planets that have not yet been detected. Utilizing these upper limits can help constrain the available parameter space in which to search for more planets, thereby allowing for efficient optimization of further search and follow-up observations with a variety of instruments.

The \tnt{} software package is a unique and powerful tool to assess exoplanetary system architectures and predict the presence and parameters of additional planets in a given system. \tnt{} was first introduced by \citet[][hereafter \citetalias{Dietrich2020}]{Dietrich2020} with an analysis of the suite of TESS multi-planet systems and expanded with multiple additional studies of individual systems \citep[][hereafter \citetalias{Dietrich2021}, \citetalias{Dietrich2022}, \citetalias{Basant2022a}]{Dietrich2021, Dietrich2022, Basant2022a}. \tnt{} uses the statistics from the Kepler population and combines that with the specific but incomplete data from individual systems to determine the statistically most likely orbital period, inclination, eccentricity, and planet size (mass/radius) for an additional planet if it exists in the system. The system with the new planet is tested with a dynamical stability criterion to see if it is dynamically stable over long periods \citep[e.g., \citetalias{Dietrich2022},][]{Dietrich2024a}. Planetary systems are also shown with their habitable zones overlaid as a function of the age of the system, showing if and/or how long planets and additional candidates would be in the habitable zone \citep[][]{Basant2022b}.

My approach is to include observational upper limits to provide a new dimension for \tnt{} to provide parameter constraints for additional planets in exoplanet systems. This paper is organized as follows. The procedural updates to \tnt{} to include the observational upper limits are detailed in Section~\ref{sec:limits}. In Section~\ref{sec:results} I apply the new version of \tnt{} to the previously-analyzed systems from \citetalias{Dietrich2021}, \citetalias{Dietrich2022}, and \citetalias{Basant2022a}, as well as for additional multi- and single-planet systems with given upper limits. I discuss my results for the observational upper limits and the expanded parameter space available for predictions in Section~\ref{sec:discussion}. Finally, in Section~\ref{sec:summary} I summarize my results and conclusions.

\section{Observational Upper Limits} \label{sec:limits}

Upper limits on detections provide constraints on the presence or lack thereof for additional planets in a given system, normally defined as a limiting size for the additional planet based on what its observable signatures would be. Note that in the context of \tnt{}, these limits do not add likelihood for additional planets outside the parameter ranges of the known planets or planet candidates, but do restrict areas in likelihood space for additional planets within the constraints of the observations. I note that these limits are dependent both on signal amplitude and timespan of observations. While all the results mentioned in this manuscript are focused on planets with orbital periods shorter than the data timespan, this technique breaks down and is not applicable for planets with longer orbital periods. While some observational surveys are structured in a way to be uniform and robust, different surveys are almost never uniformly managed, and also each planetary system has similar but unique architectural parameters. Therefore, observational upper limits won't have a standard statistical distribution across all of the known planetary systems, and usually need to be included on a per-system basis.

Here I use upper limits from RVs, transits, and TTVs as these are currently the most prevalent and sensitive, with the long baselines of RV observations for over three decades and the very high precision of Kepler and TESS photometry. Direct imaging limits and astrometry limits could be included in the future, but are not currently included. Direct imaging is usually used at large separations for a planet population nominally different than the Kepler population, whereas astrometry limits from future Gaia data releases will be the standard. RV and transit sensitivities are usually calculated over a range in planetary orbital period, while TTV sensitivities are measured for each known planet. Here I describe how the observational upper limits were included in \tnt{} for each of these types of limits, in conjunction with the already known data, the Kepler population statistics, and the mutual Hill radius dynamical stability criterion from \citetalias[][]{Dietrich2022} and improved by \citet[][]{Dietrich2024a}.

\subsection{RVs} \label{subsec:ul_rv}

RV upper limits are nominally given in some statement similar to the following: ``We place an upper limit on the minimum mass for additional planets of XX $M_\oplus$ (or $M_{Jup}$) for the orbital period range of YY-ZZ days," or ``We place an upper limit on the RV semi-amplitude for additional planets at XX m/s for the orbital period range of YY-ZZ days." RV upper limits are also dependent on whether or not any or all of the previously found planet candidates have had their putative signal subtracted. This is due to being based on the residuals after the chosen signals have been extracted; see e.g., \citet[][]{Cretignier2023} for an example of how an RV upper limit used later in this analysis is determined. For \tnt{}'s analysis, I translate that into a constraint on additional planets with the following steps:

\begin{itemize}
    \item I model the orbital period, planet radius (to translate to mass), orbital inclination, and orbital eccentricity distribution for an additional planet as in \citetalias{Dietrich2020} and \citetalias{Dietrich2022}. All of these parameters, as well as the stellar mass, factor into the RV semi-amplitude formula \citep[see e.g.,][]{Perryman2011}.
    \item For every Monte Carlo planet injection, I check the expected orbital period to see if it is within the range for an RV upper limit. If so, I calculate the expected mass of the injected planet from the planet radius model distribution and a M-R relationship.
    \item I calculate the RV semi-amplitude given the predicted mass, period, and inclination of the injected planet. If the expected injected planet RV semi-amplitude or mass is larger than the given limit, I reject the injection. If not, I accept it.
\end{itemize}

\subsection{Transits} \label{subsec:ul_ts}

Transit upper limits are usually given in a similar way to RV upper limits, except the observable parameter that is being limited is the transit depth instead of the RV semi-amplitude: ``We place an upper limit on the transit depth for additional planets of XX ppm for an orbital period range of YY-ZZ days." This then can be translated into an upper limit on the planet radius via the flux ratio equation, although this does not fully take into account the dependence on the impact parameter - a larger planet at a high impact parameter can have the same transit depth as a smaller planet at a low impact parameter. In addition, longer-period transiting planets have longer transit durations and longer times between transits, so the sensitivity and SNR for transiting planets decreases with increasing orbital period. I follow an analogous procedure to the RV upper limits in order to include transit upper limits to the \tnt{} framework:

\begin{itemize}
    \item I model the orbital period, planet radius, orbital inclination, and orbital eccentricity distribution for an additional planet as in \citetalias{Dietrich2020} and \citetalias{Dietrich2022}.
    \item For every Monte Carlo planet injection, I check the expected orbital period to see if it is within the range for a transit upper limit. If so, I determine if the injected planet will transit given its expected orbital period and inclination.
    \item If the injected planet is expected to transit and the expected transit depth is larger than the upper limit, I reject the injection. If either of these is not true, I accept the injection.
\end{itemize}

\subsection{TTVs} \label{subsec:ul_ttv}

The presence or limit on TTVs provides a separate and different constraint than the previous two limits. TTVs are measured for each planet, and an upper limit can be given on the presence of TTVs if a clear pattern is not discernible. This is usually given in the form ``We find TTVs [or place an upper limit on the presence of TTVs] for planet XX at a value of YY minutes." I note that this analysis is restricted to resonance effects as they provide the strongest signals for our sensitivity and timespan. Additional effects like apsidal precession, the R{\o}mer effect, and the Applegate effect are mostly only significant for Jovian planets and/or over long timescales, and would likely produce signals with amplitudes $<<$ 1 second for shorter period lower mass planets, or signals with amplitudes of $\sim1$ minute over multiple years \citep[see e.g.,][]{Agol2005, Mannaday2022, Wang2024}. Thus, due to the timescales of those effects being much longer than the orbital periods of the planets we are measuring, these effects will not be considered. To include TTVs caused by additional planets due to resonance effects and their limits, I outline the following procedure:

\begin{itemize}
    \item I model the orbital period, planet radius, and orbital inclination distribution for an additional planet as in \citetalias{Dietrich2020} and \citetalias{Dietrich2022}.
    \item For every Monte Carlo planet injection, I model the TTVs for each planet in the new system. This can currently be done by two different methods implemented in the code.\\
    
    Note that both of these methods, when the data is unavailable, use random parameters for the arguments of periastron and true anomalies. For multi-planet systems with many transiting planets, the longitude of ascending node is expected to be similar for all the planets and could therefore be chosen to be any specific value \citep[e.g.][]{Grimm2018}. However, in general we cannot assume that for planetary systems where none, one, or a minority of the planets transit. The longitude of ascending node, argument of periastron, and the true anomaly (or alternatively the time of periastron, which is the time when the true anomaly is 0) are not known for $\sim$75-90\% of planets in the NASA Exoplanet Archive \citep[][]{ExoplanetArchive}\footnote{Accessed most recently on 2024-06-27 at 12:49, returning 5539 rows.}. In these cases, I use random values between 0 and 360 degrees for the angles, or between 0 days and the orbital period of the planet for the time of periastron.

    \begin{itemize}
        \item The first method to model TTVs is to run an N-body integration with REBOUND \citep[][]{Rein2012, Rein2019}. With sufficient time resolution, TTVs on the order of a couple minutes can be measured, albeit with a longer runtime. The integrations are run for 1000 transits of the innermost planet with dynamical timesteps starting at 20\% of the innermost planet period and decreasing to find the midpoint of the transit.\\
        \item The second method to model TTVs is to utilize the TTVFaster algorithm from \citet[][]{Agol2016a} to quickly but accurately measure the expected TTVs from the Keplerian orbital parameters for every planet from a given system. The results I report in Section~\ref{sec:results} use TTVFaster with known or random orbital angles as described below.
    \end{itemize}
    
    \item These resulting TTV values are highly dependent on the orbital angles, and so for the large majority of systems that do not have known orbital angles, I instead provide the entire array of expected TTVs produced by the new system with the injected planet across the different orbital angles.
    \item For planets with known orbital angles, if the given TTVs are above the limit I reject the injection, otherwise I accept it. For planets with unknown orbital angles, I draw a random number between 0 and 1. If the TTV limit for a planet is below the corresponding percentile of the expected TTV distribution for that planet, I reject the injection. If it is higher, I accept the injection.
\end{itemize}

\section{Results} \label{sec:results}

Here I show the results of the new \tnt{} analyses incorporating the additional observational limits on three individual systems which have been previously analyzed by \tnt{}: \ensuremath{\tau} Ceti \citepalias{Dietrich2021}, HD 219134 \citepalias{Dietrich2022}, and HD 20794 \citepalias{Basant2022a}. The planets and planet candidates in these systems are mostly RV detections, although HD 219134 has two planets that also transit. Additional work on a larger variety of systems is discussed in Section~\ref{subsec:disc_ns}.

\subsection{\ensuremath{\tau} Ceti} \label{subsec:res_prev_tauCeti}

\ensuremath{\tau} Ceti, the closest single star similar to the Sun \citep[spectral class G8V, distance 3.65 pc;][]{Hall2004, Teixeira2009}, is reported to have between four and eight planets/planet candidates, seven of which are RV signals of likely small planets in the inner system \citep[periods between 14 and 636 days, $m \sin i < 4 M_\oplus$;][]{Tuomi2013, Feng2017b} and one cold Jupiter inferred via astrometric excess \citep[$m \sim 1-2 M_{Jup}$, $a = 3-20$ au;][]{Kervella2019}. $\tau$ Ceti is also known to have low stellar variability of $\sim$1 m/s and is thus a good RV standard star for high- and extreme-precision RV instruments like HARPS, EXPRES, and NEID \citep[e.g.,][]{Pepe2011, Lovis2011, Tuomi2013}. The RV signals of the inner planets/candidates are all sub-1 m/s and even approaching 0.5 m/s, with the possibility remaining that all of the signals are of unconfirmed candidates and may not be bona fide planets \citep[e.g.,][]{Pepe2011, Cretignier2021, Cretignier2023}. Here I assume the planets found by \citet[][]{Feng2017b} are true planets and the rest are unconfirmed candidates, as this is the current status of the system in the NASA Exoplanet Archive.

The inclination of the planets is unknown, so the minimum mass is the only indication of size I have for the RV planets. \ensuremath{\tau} Ceti is known to have a debris disk which is best-fit by models at $i = 35^\circ \pm 10^\circ$ \cite[][]{Lawler2014, MacGregor2016}. If the planets/candidates are aligned with the disk, their mass would be roughly twice the minimum mass and very likely stable in the current configuration. However, \citet[][]{Korolik2023} used a combination of RV and interferometric imaging data to find that \ensuremath{\tau} Ceti is likely rotating very near to pole-on from my point of view, with a rotation axis of $7^\circ \pm 7^\circ$. This is a significant misalignment with the debris disk, and also greatly increases the mass of the exoplanets if they are more aligned with the stellar rotation axis. With these higher masses and the best-fit eccentricities from \citet[][]{Feng2017b}, the system would more likely be unstable. Either an increase of inclination away from pole-on or a decrease in the expected eccentricities \citep[probably a result of instrument noise and fitting bias, as suggested in e.g.,][]{Feng2017a, Feng2017b} would allow for the current configuration of planets and candidates to be dynamically stable over the billions of years the system has existed.

\begin{figure*}[ht]
    \centering
    \includegraphics[width=0.49\linewidth]{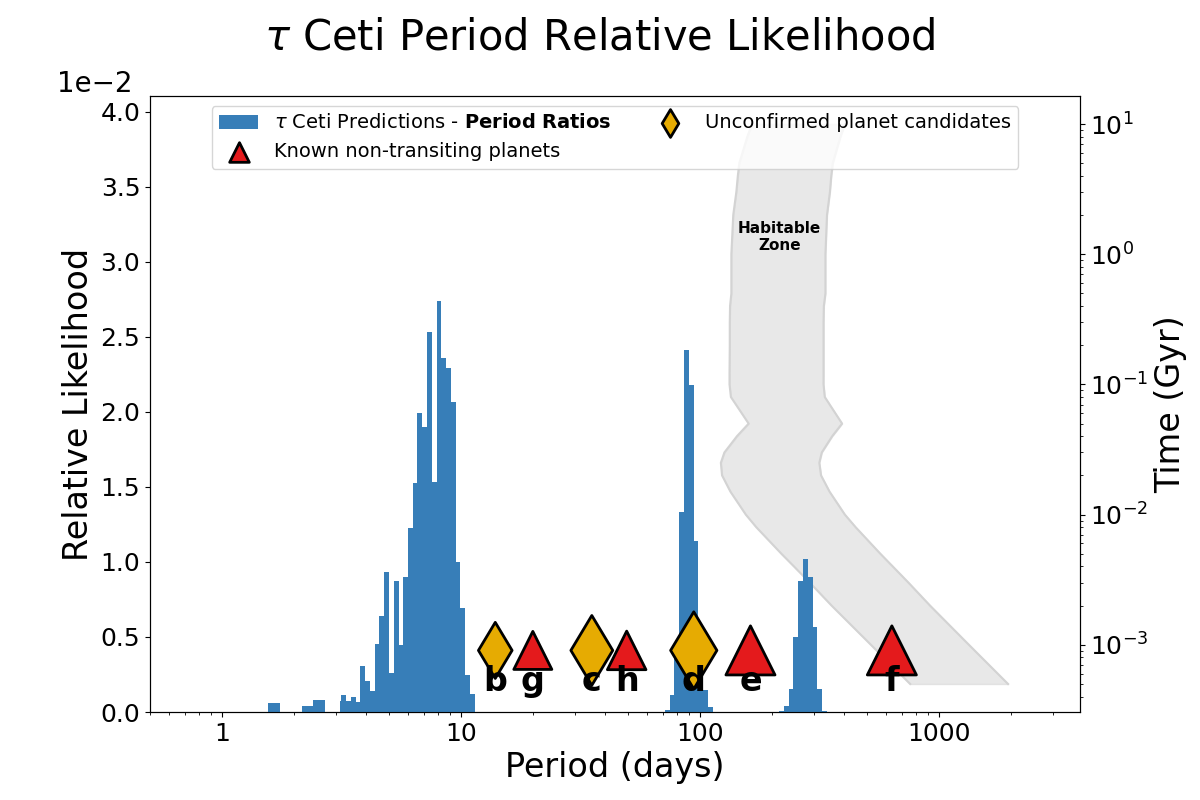}
    \includegraphics[width=0.49\linewidth]{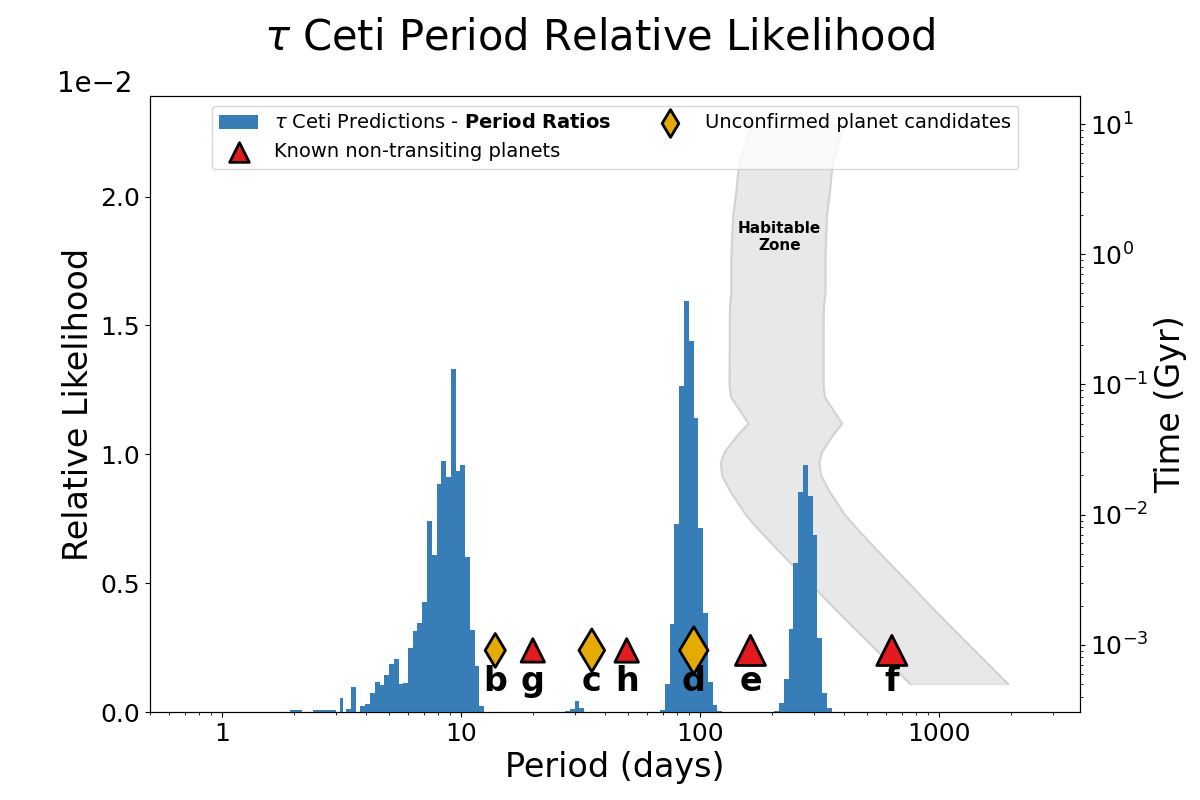}\\
    \includegraphics[width=0.49\linewidth]{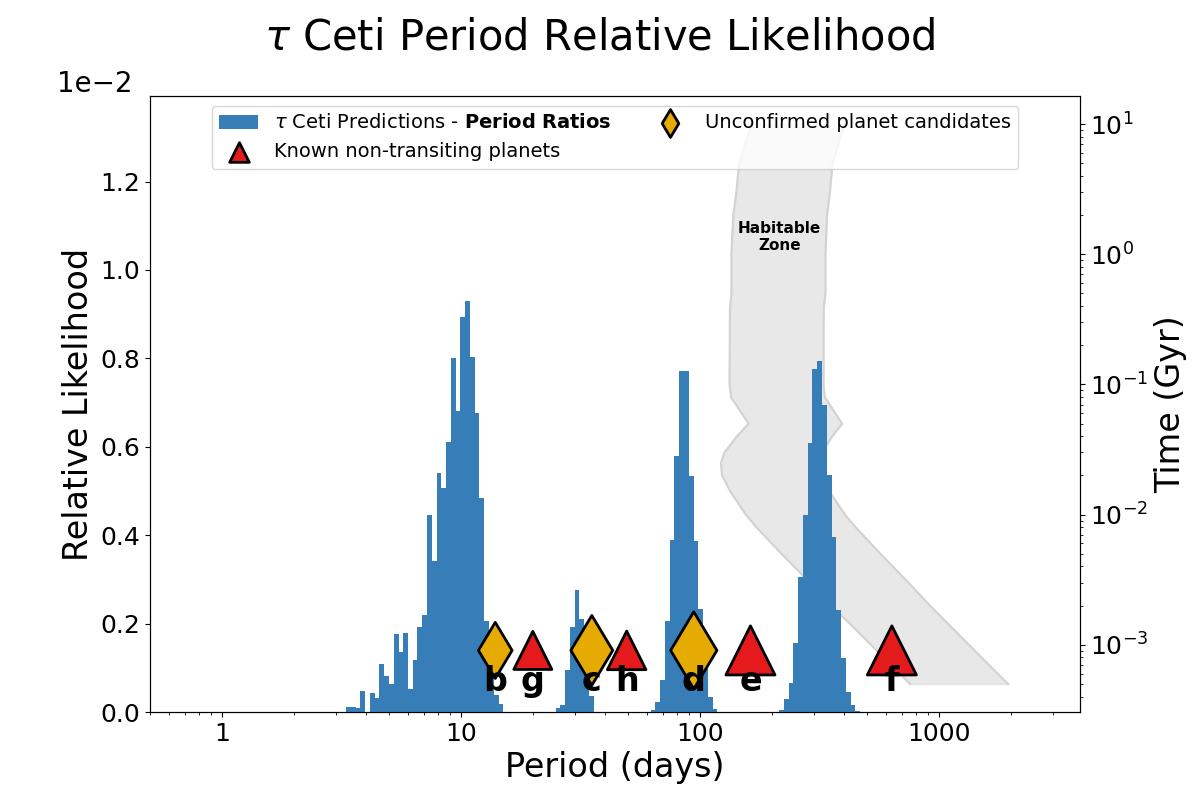}
    \includegraphics[width=0.49\linewidth]{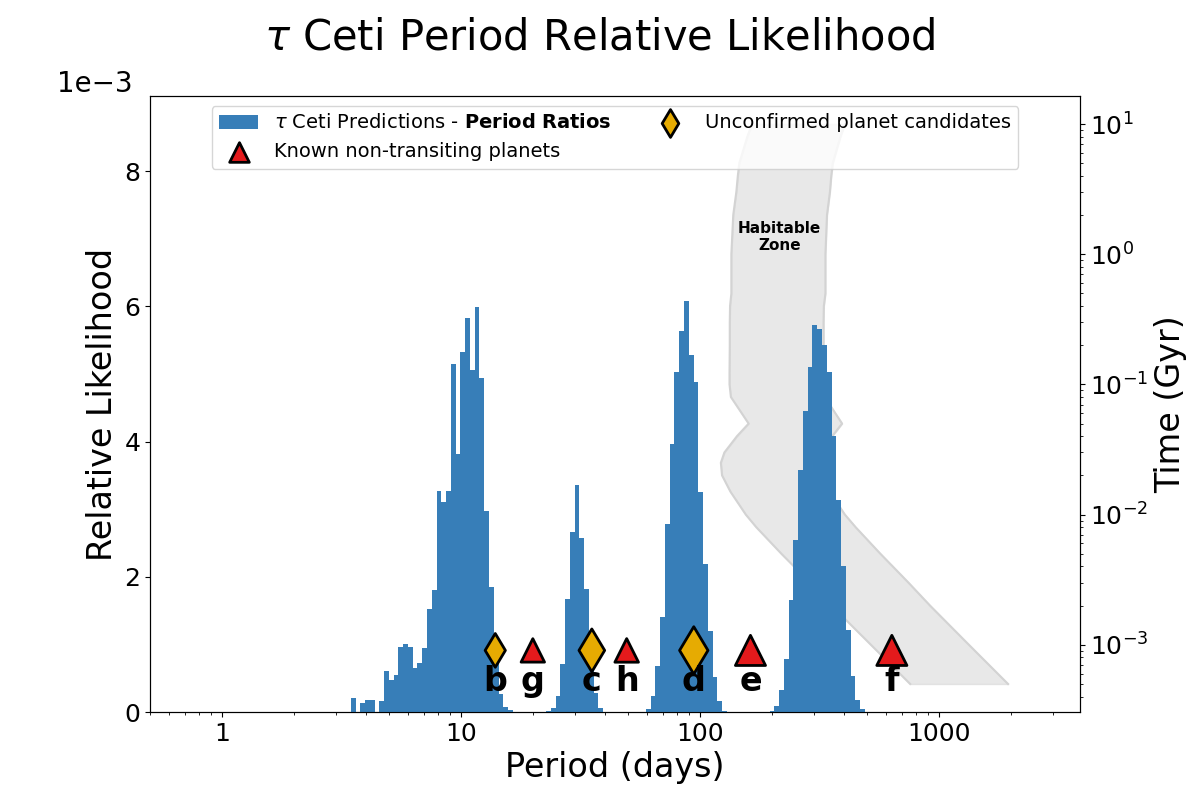}\\
    \caption{\tnt{} predictions in orbital period space for tau Ceti where the planets have eccentric (top) and circular (bottom) orbits, as well as inclinations of 7$^\circ$ (left) and 35$^\circ$ (right). The eccentricity of the orbits plays a stronger role than the inclination (and therefore mass) of the planets for decreasing the stability of the system with an additional planet. As in \citet{Basant2022b}, the habitable zone is plotted here as a function of age of the system (shown on the right vertical axis) to highlight the effect of the host star evolution on the habitable zone location.}
    \label{fig:tauCeti}
\end{figure*}

A previous analysis of \ensuremath{\tau} Ceti in \citetalias{Dietrich2021} using \tnt{} had some notably different assumptions. In particular, the planets were assumed to orbit aligned with the debris disk, as many systems have found such planetary orbit alignments \citep[see e.g.,][]{Apai2015, Plavchan2020}, and the orbits were assumed to be circular, which was reasonable due to the correlation between high eccentricity and instrumental red noise mentioned by \citet[][]{Feng2017b}. This led to high likelihood of the planet candidates b, c, and d from \citet[][]{Tuomi2013} being genuine, as the analysis showed peaks of probability where an additional planet in the system would be stable. However, when the planets are allowed to have non-zero eccentricities and lower (i.e., more face-on) inclinations, planet candidates c and d become much less likely to be genuine, and the predicted additional candidate in the habitable zone (as defined by \citet[][]{Kopparapu2013,Kopparapu2014}) of \ensuremath{\tau} Ceti in \citetalias{Dietrich2021} also becomes much less likely to exist. A comparison of the new \tnt{} predictions for circular and non-circular orbits, as well as the two different measured system inclination proxies, is shown in Figure~\ref{fig:tauCeti}.

For including upper limits, I first tested a generic 1 m/s upper limit on the RV signal of any additional planet with orbital periods less than 300 days or greater than 450 days.
The 1 m/s upper limit is given by the stellar variability, as seen through more than a decade of RV observations, and is roughly twice the strength of the signals seen for the smallest planets and candidates in the HARPS data \citep[][]{Feng2017b}, as well as roughly three times larger than the limit for the current EPRV instruments \citep[e.g.,][]{Jurgenson2016, Halverson2016}. For the given inclination of the disk and circular orbits, this limit corresponds to $\sim 5 M_\oplus$ at 10 days but $\sim 17 M_\oplus$ at 400 days; for comparison, the most massive planets discovered by RV so far would have masses $\sim 7 M_\oplus$ at the disk inclination. At the inclination of the stellar rotation axis with eccentricities near those measured by \citet[][]{Feng2017b}, however, the RV limit of 1 m/s at orbital periods of 10 and 400 days would correspond to planet mass upper limits of $\sim 23 M_\oplus$ and $\sim 80 M_\oplus$.

\begin{figure*}[ht]
    \centering
    \includegraphics[width=0.49\linewidth]{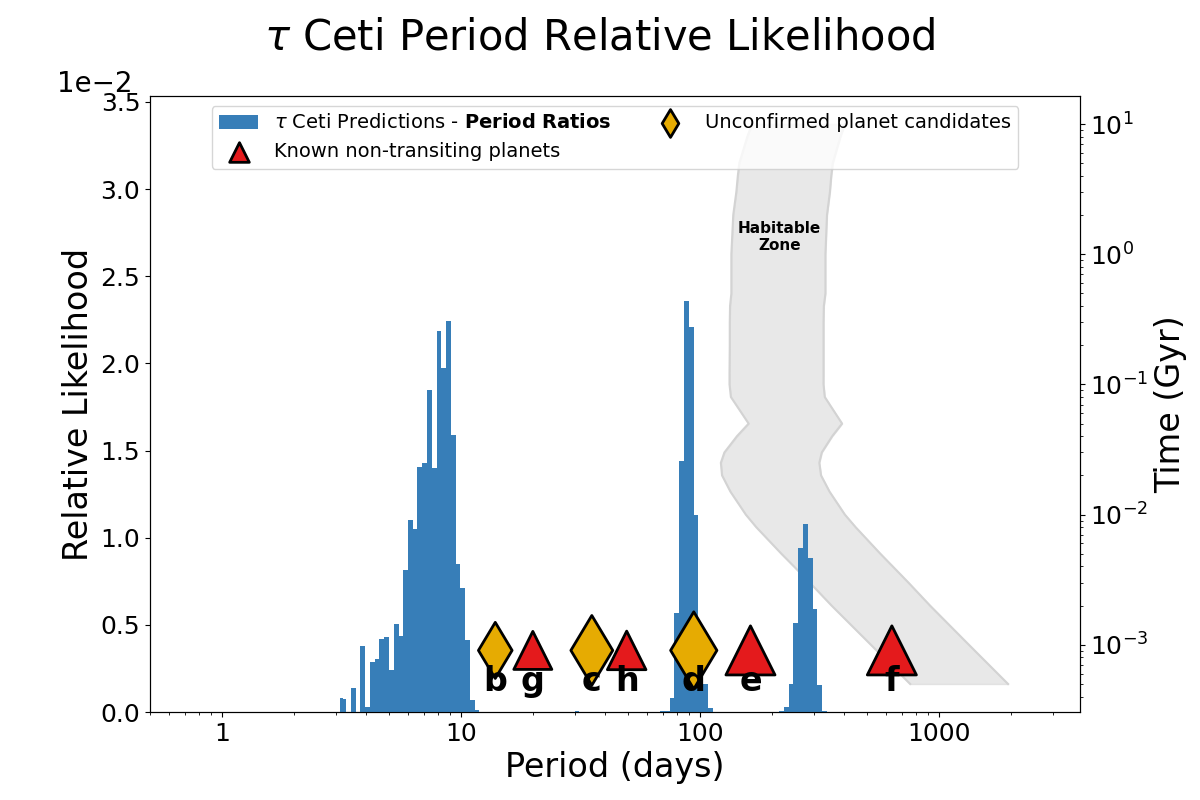}
    \includegraphics[width=0.49\linewidth]{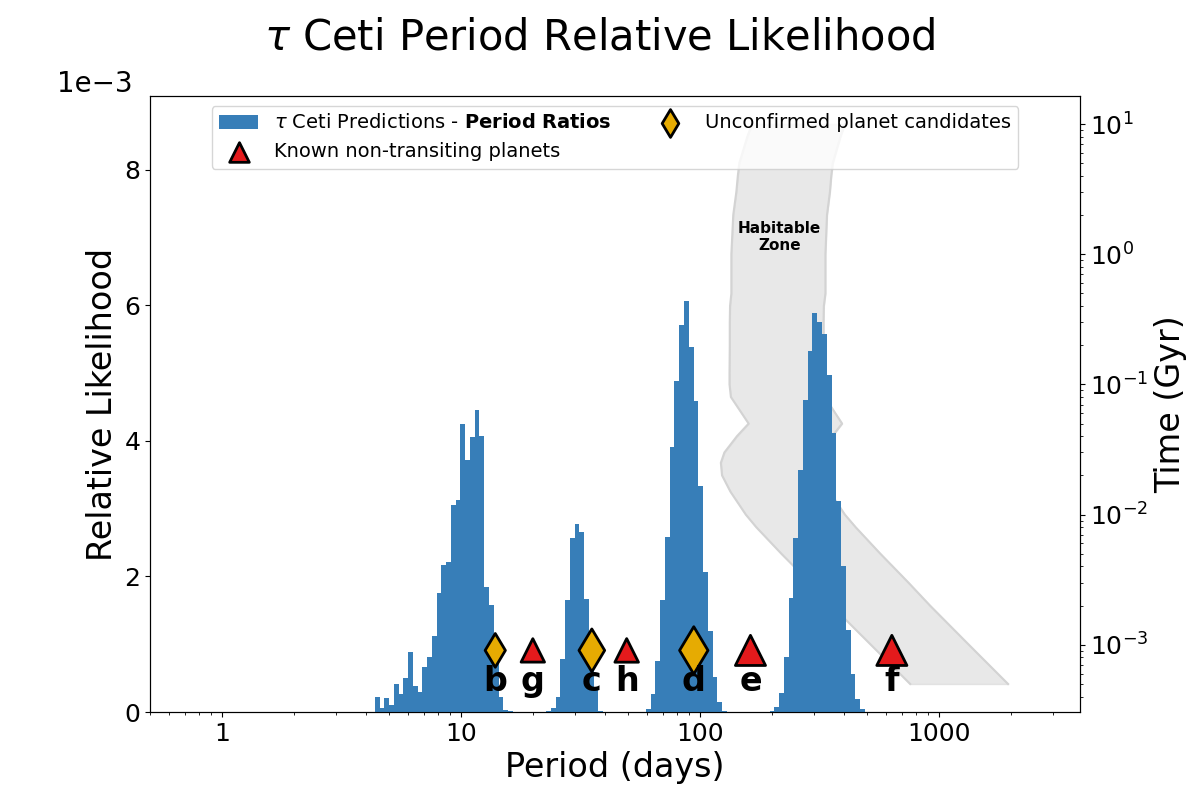}\\
    \includegraphics[width=0.49\linewidth]{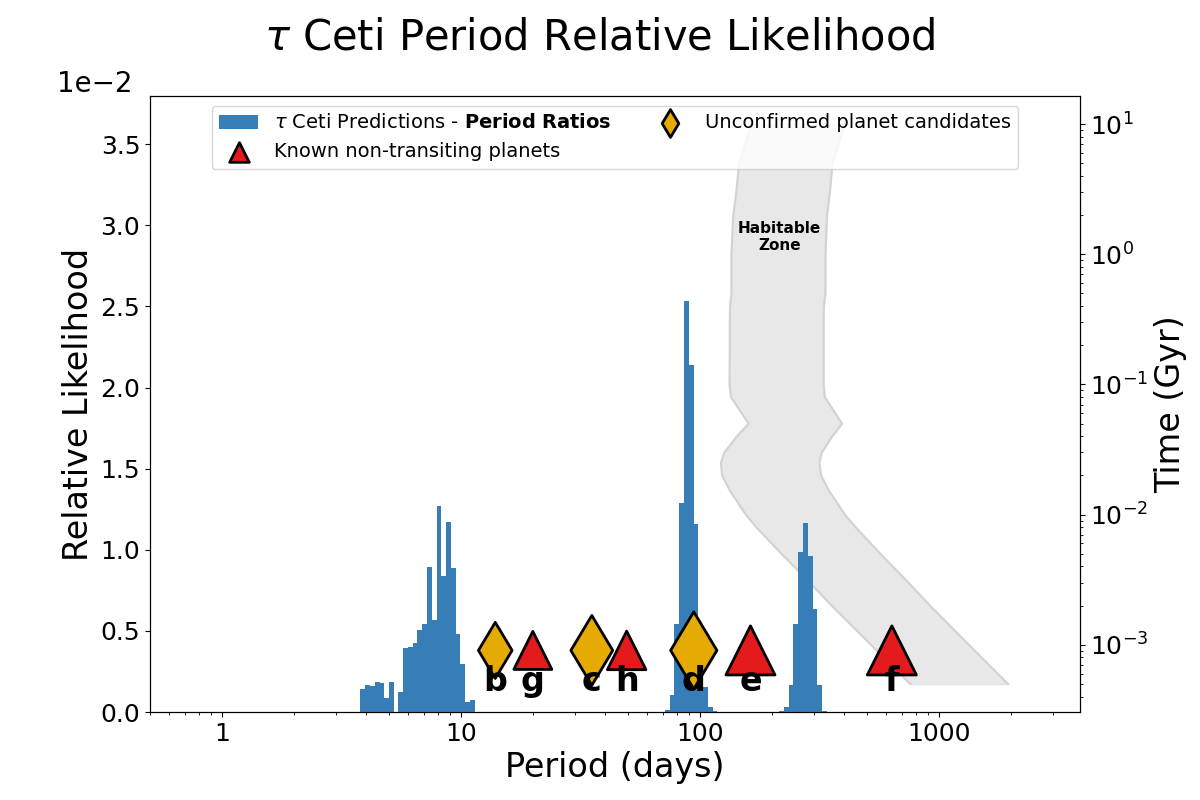}
    \includegraphics[width=0.49\linewidth]{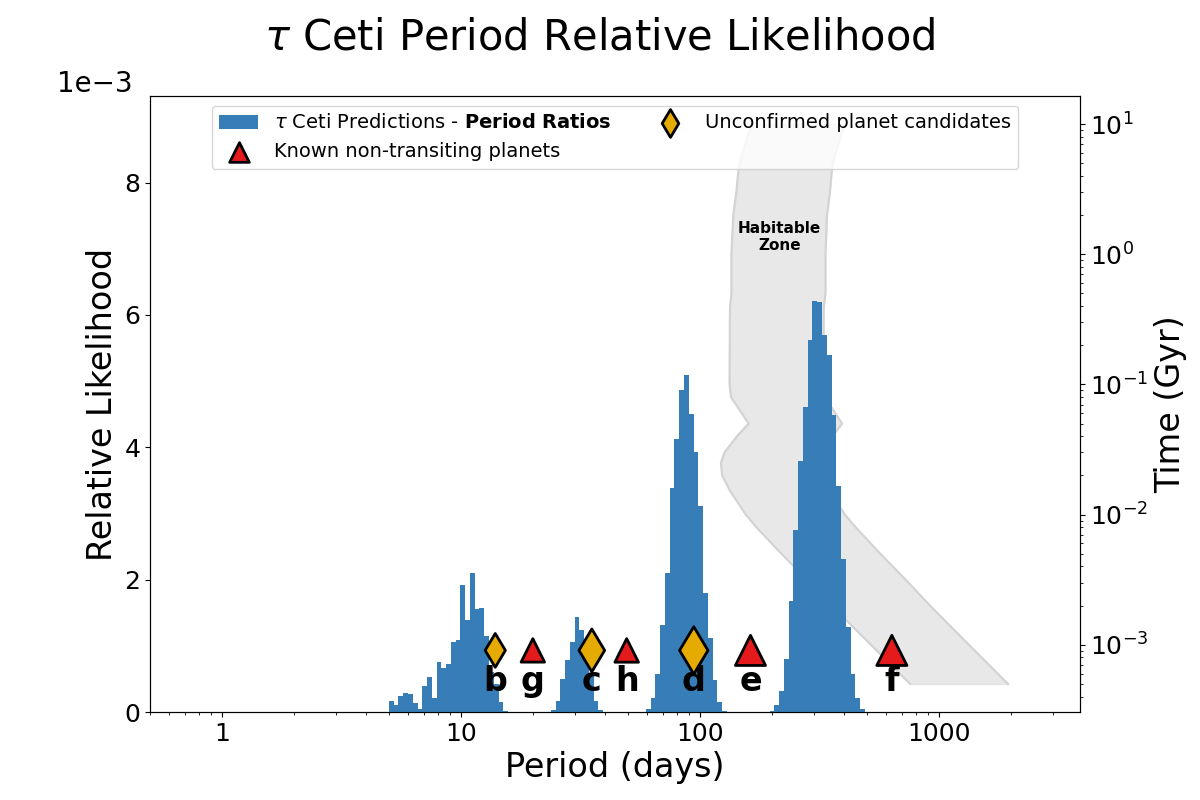}\\
    \caption{\tnt{} predictions in orbital period space for tau Ceti where the RV limit is 1 m/s (top) and 0.5 m/s (bottom) for planets with eccentric orbits and inclinations of 7$^\circ$ (left) as well as circular orbits with inclinations of 35$^\circ$ (right). The 1 m/s limits do not significantly change the predicted values from those made without limits, while the 0.5 m/s limits does cause a lower likelihood of a planet existing near a 10 day orbital period, near the location of planet candidate b from \citet[][]{Tuomi2013}. Habitable zones are shown like in Figure~\ref{fig:tauCeti}}.
    \label{fig:tauCetiRVlims}
\end{figure*}

\begin{figure}[ht]
    \centering
    \includegraphics[width=\columnwidth]{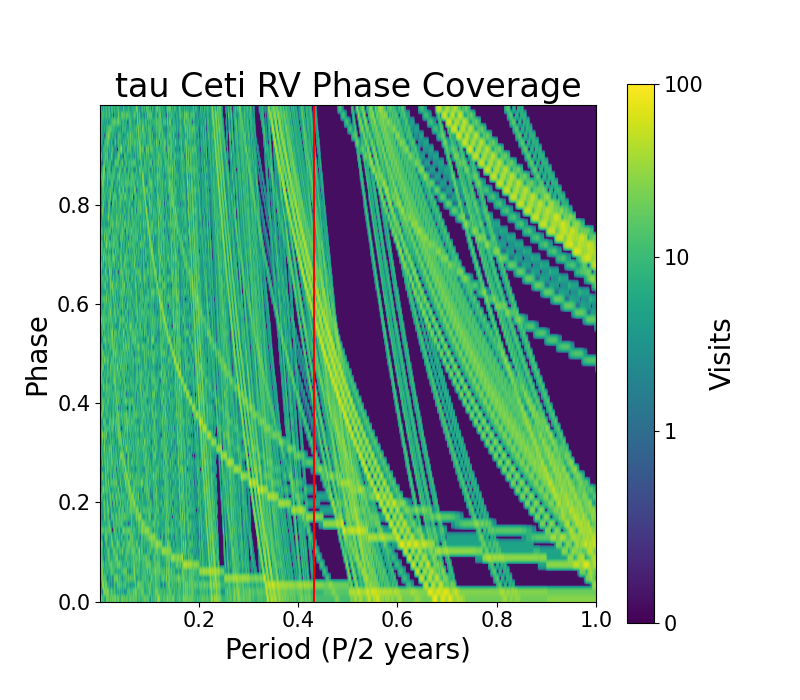}
    \caption{The period phase coverage for tau Ceti with RV observations from NEID and EXPRES, shown in number of visits to a specific phase per orbital period (expressed as a fraction compared to 2 years). The tentative signal from \citet[][]{Tuomi2013} is shown as a red line. With the number of visits above 10 for most of the phases at this period, I find it unlikely the signal is a real planet.}
    \label{fig:tauCetiRV}
\end{figure}

The comparison between these upper limits is shown in Figure~\ref{fig:tauCetiRVlims}. I find that adding in a RV upper limit of 1 m/s does not significantly affect the predictions from \tnt{}. When lowering the limit to 0.5 m/s, this decreases the probability of finding a planet interior to planet g, as similar mass planets have higher RV semi-amplitudes at shorter periods and so an additional planet at 10 days (close to where planet candidate b was first reported by \citet[][]{Tuomi2013}) would therefore be more likely to be above the upper limit and be rejected by the Monte Carlo iterations. In addition, \ensuremath{\tau} Ceti has been observed by NEID (Bender, private communication) and EXPRES \citep[][]{Korolik2023}for multiple seasons, and the combined phase coverage of the two instruments is very comprehensive over periods ranging up to two years (see Figure~\ref{fig:tauCetiRV}), suggesting that any additional planets with an RV signal above 0.5 m/s would likely have been caught. However, gaps due to the 1-year orbital period of Earth could occur around the half phases of the period, only showing the quarter phases where the RV would remain relatively constant over different phase cycles.

\begin{figure*}[ht]
    \centering
    \includegraphics[width=0.49\linewidth]{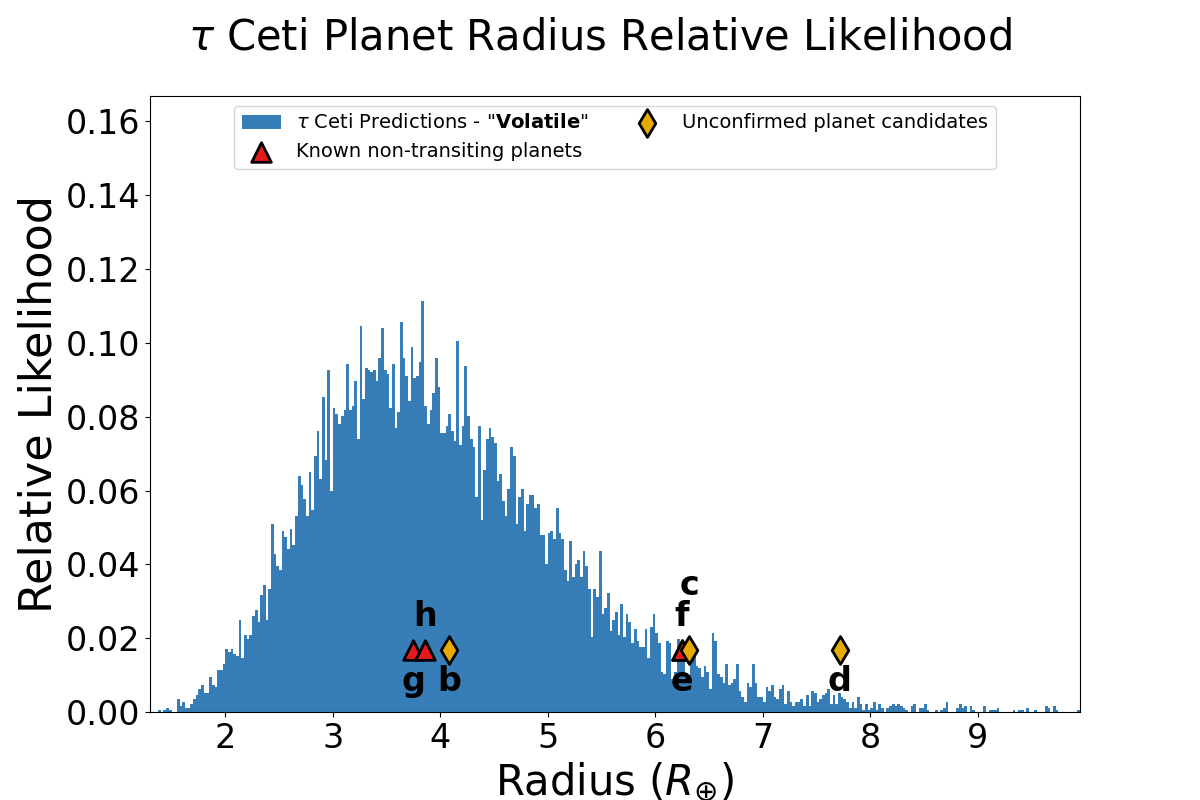}
    \includegraphics[width=0.49\linewidth]{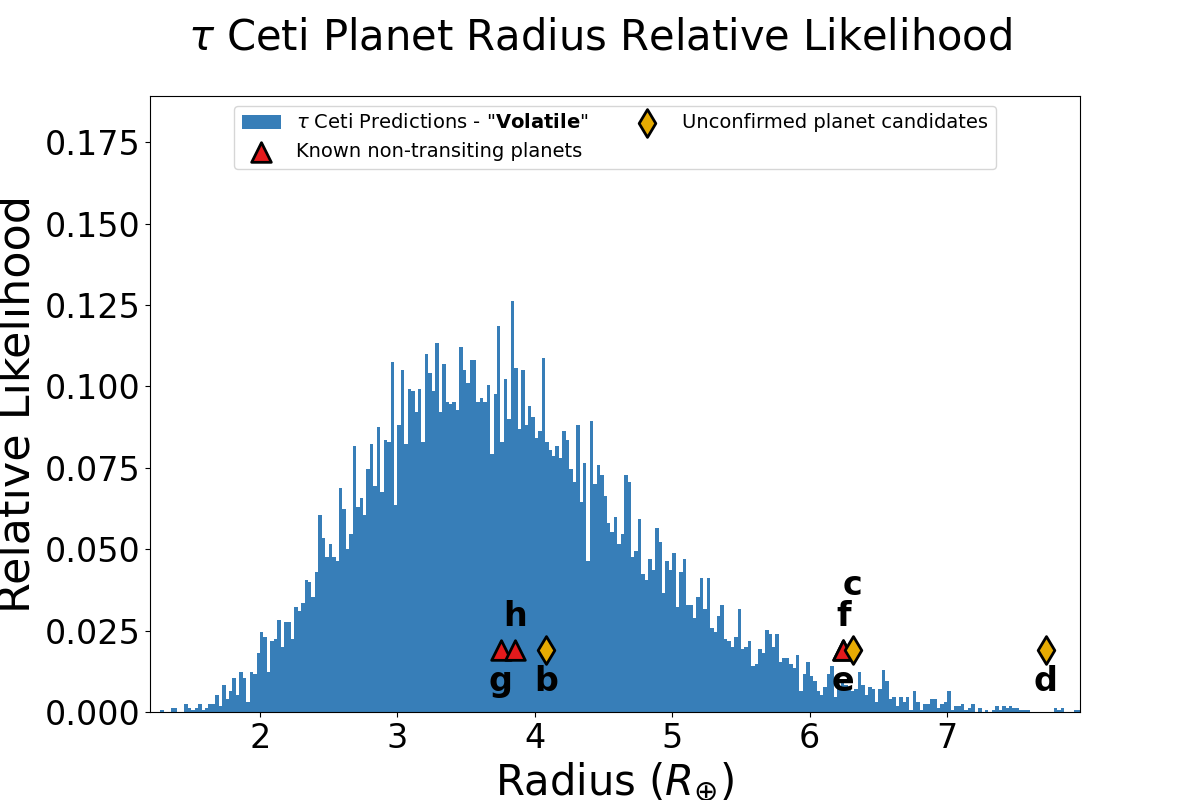}\\
    \caption{The predicted planet radius distribution as calculated from an assumed planet orbital inclination similar to the stellar rotation axis of $7^\circ$ and a volatile-rich atmosphere for the mass-radius relationship from \citet[][]{Otegi2020}, with no RV limits enforced (left) and a 0.5 m/s RV limit enforced (right). There is no significant difference between the relative likelihoods for the different planet radii.}
    \label{fig:tauCetisize}
\end{figure*}

The effect of the RV limits is mostly seen on the orbital period likelihoods. The inner system planets and candidates would be sub- to super-Neptunes based on the inclination, but with such a high inclination the $\sin i$ term lowers the semi-amplitude, and thus allows for the standard range of predicted planet sizes. This is nominally given in planet radius because the Kepler statistics are in planet radius. For planets only with measured minimum masses, we select an inclination measure, which is either chosen randomly from an isotopic distribution in $\cos(i)$ or using values given in the literature, as in the case of \ensuremath{\tau} Ceti of $7^\circ$ or $35^\circ$ based on the debris disk inclination or the stellar rotation axis. Thus, with a measured or calculated mass, we use a mass-radius relationship to convert to radius and back as in \citetalias{Dietrich2020} and \citetalias{Dietrich2021}. In this case we use the broken power-law relationship from \citet[][]{Otegi2020} and assume a volatile-rich atmosphere as most of the planet candidates orbit far enough away that they could retain a gas envelope \citep[see e.g.,][ for differences between sub-Neptunes and super-Earths as a function of orbital period]{Bergsten2022}. I show the comparison between no RV limits and 0.5 m/s RV limits for the $7^\circ$ inclination and eccentric orbit case in Figure~\ref{fig:tauCetisize} and note that there is little difference between the likelihoods.

In summary, for \ensuremath{\tau} Ceti I find that the moderately-high eccentricities given for the planets, in conjunction with potentially a more face-on orbital inclination, precludes high likelihoods for planet candidates b, c, and d as compared to the circular and less face-on analysis from \citetalias{Dietrich2021}. When including an RV upper limit of 0.5 m/s (a realistic expectation based on residuals from HARPS, NEID, and EXPRES observations), the probability of finding a planet interior to planet g or between planets g and h decreases as well. However, there is still some likelihood between planets h and e (more likely in the eccentric and more face-on configurations), as well as planets e and f (more likely in the circular and less face-on configurations), that is allowed with both the updated dynamical constraints and the new observational constraints. If the predicted planet were to lie between e and f, it would be located in the habitable zone and near a 1-year orbital period, making it more difficult to be observed via RV from the Earth.

\subsection{HD 219134} \label{subsec:res_prev_HD219134}

HD 219134 is an early K-dwarf \citep[$M \sim 0.8 M_\odot$,][]{Boyajian2012} star at $6.542 \pm 0.004$ pc from the Sun \citep[][]{Gaia2023}. It is the closest star with transiting planets and also the closest star with a 6-planet system \citep[][]{Motalebi2015,Vogt2015,Gillon2017a}. The inner two planets transit the host star while the outer four planets do not; multiple studies have placed upper limits on the orbital inclination of planets f and d due to these non-detections of transits \citep[][]{Gillon2017a,Seager2021}. The 22-day period planet has been deemed controversial due to it being very near the observed rotation period of the star through its activity cycle \citep[][]{Johnson2016}, but the signal itself was seen in three different studies \citep[][]{Vogt2015,Gillon2017a,Rosenthal2021} so it is still included as a confirmed planet here.

Planet g was originally found at an orbital period of 94 days by \citet[][]{Vogt2015} only, and not by either \citet[][]{Motalebi2015} or \citet[][]{Gillon2017a}. Additional RV observations from the California Planet Search (CPS) and assisted by Keck and the Automated Planet Finder (APF) did not find evidence for a planet at 94 days but instead found a significant signal at 192 days \citep[][]{Hirsch2021}. The California Legacy Survey (CLS) study also in 2021, however, found that the 94-day period was correct for planet g and the 192-day period was a false positive caused by either an annual or instrumental systematic \citep[][]{Rosenthal2021}. Thus, here I include planet g at its 94-day period, which is roughly 2x as long as planet f.

\begin{figure*}[ht]
    \centering
    \includegraphics[width=0.49\linewidth]{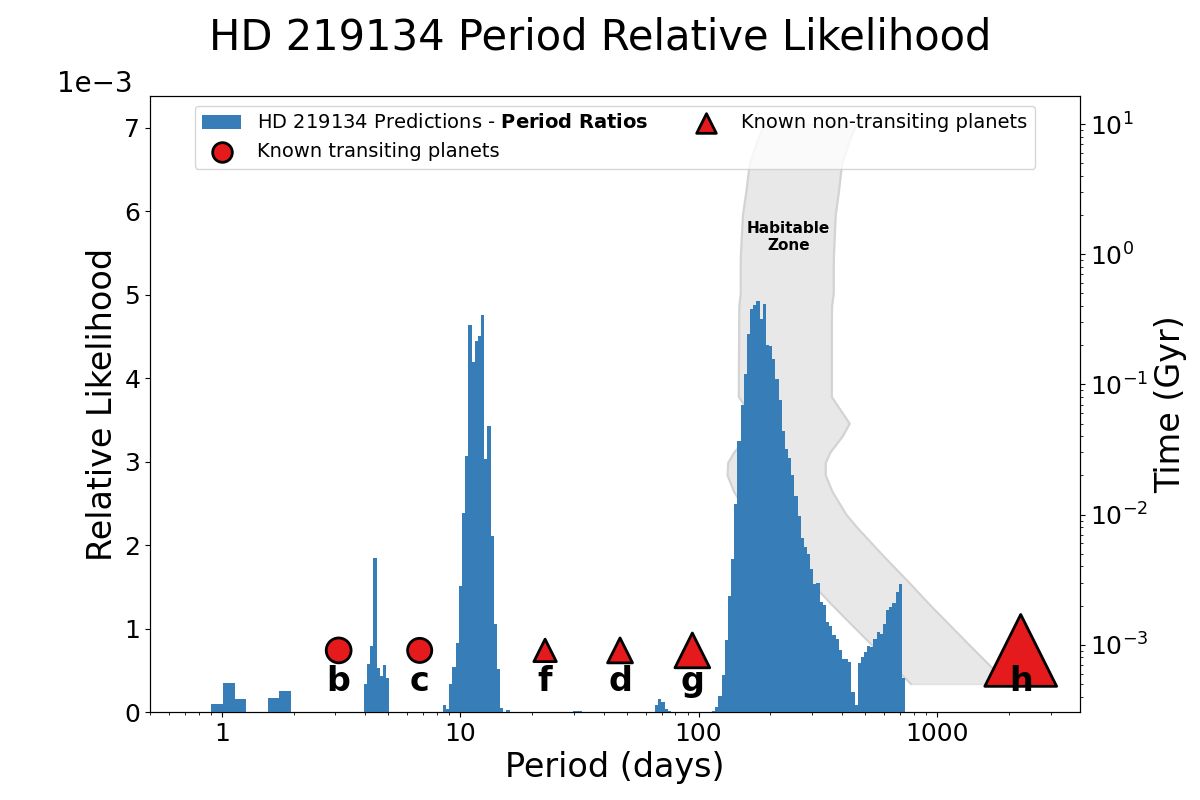}
    \includegraphics[width=0.49\linewidth]{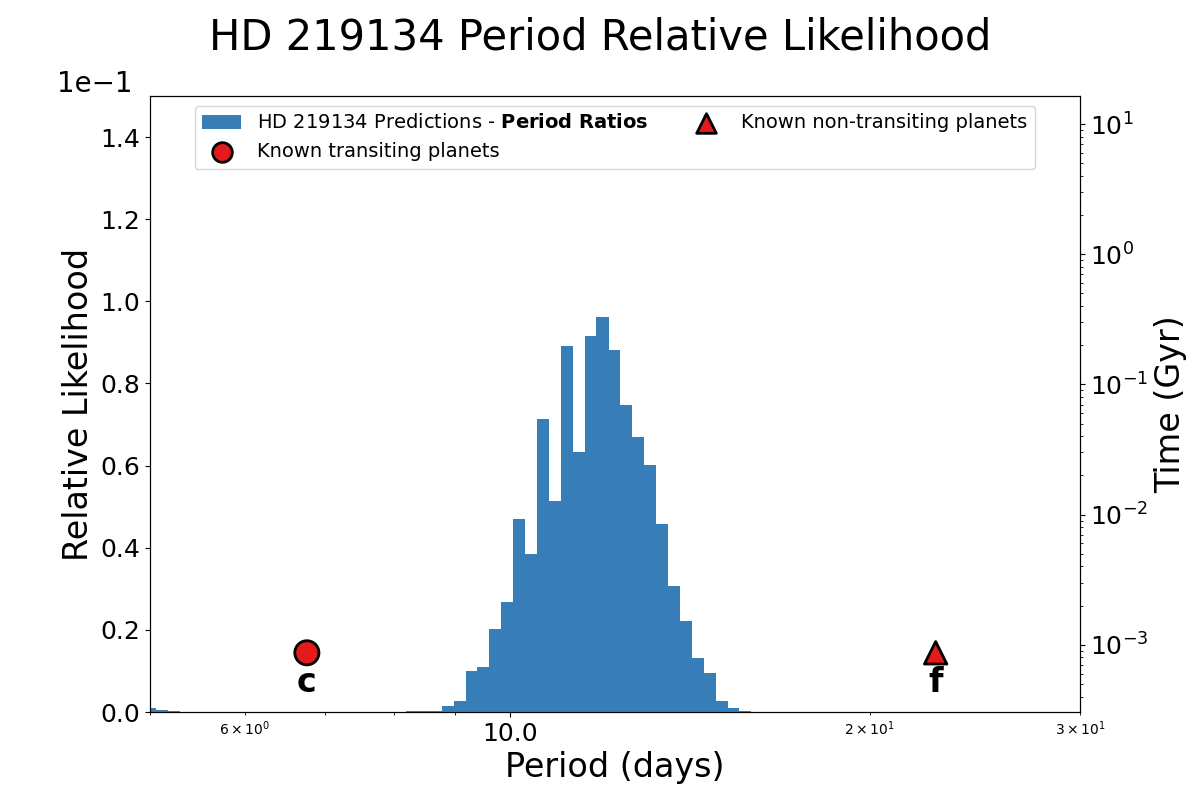}\\
    \includegraphics[width=0.49\linewidth]{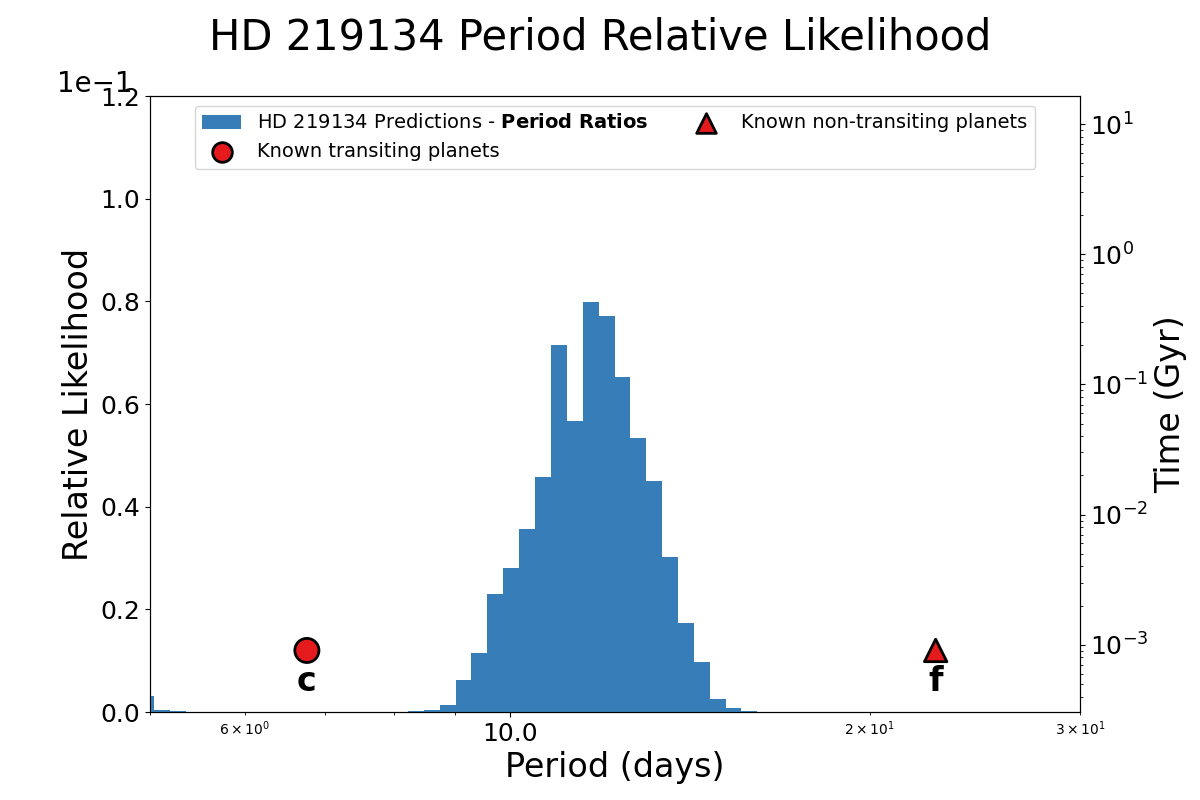}
    \includegraphics[width=0.49\linewidth]{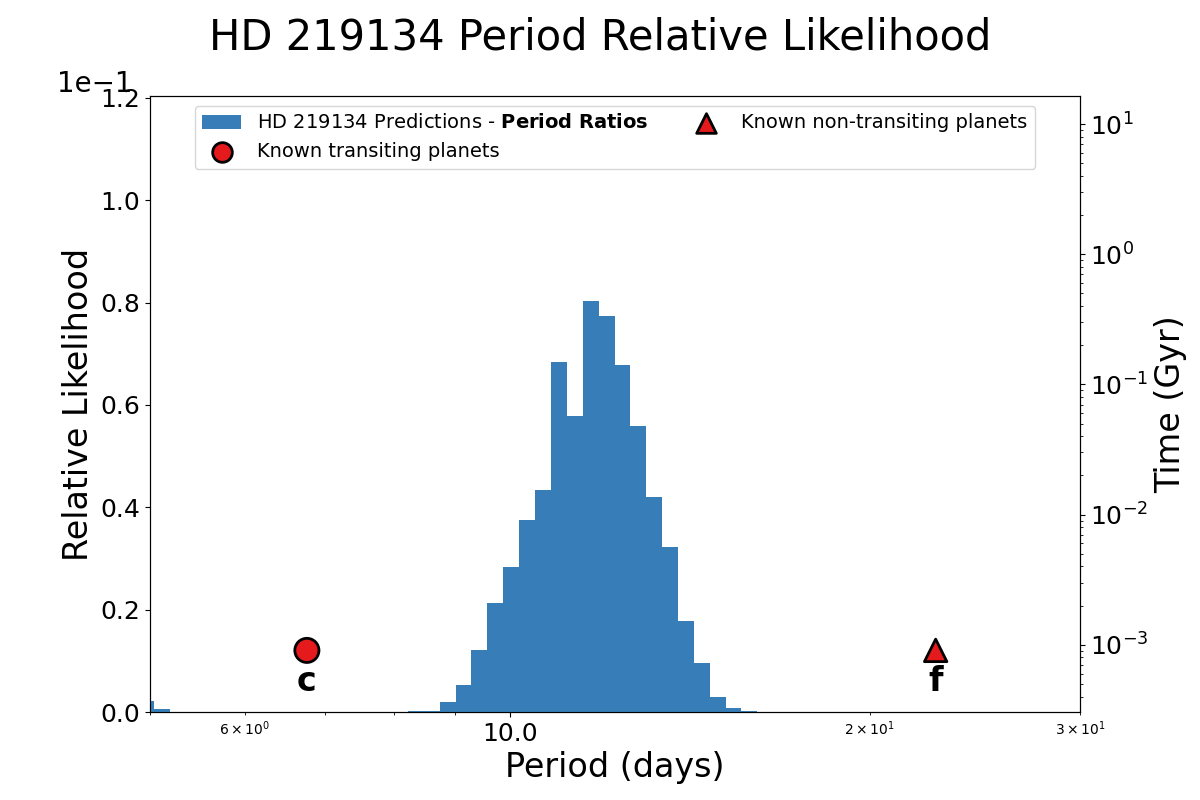}\\
    \caption{\tnt{} predictions in orbital period space for the HD 219134 system. Top left: Full predictions for 0.5--730 days; the bimodal peak with the sharp cutoff between planet g and h is due to the long period ratio between those two planets, where it is more likely to find a planet close to the Kepler ratio with g or with h, but not necessarily both. Top right: zoom in on peak around 12.4 days. Bottom left: zoom region with transit upper limit equal to smallest found planet. Bottom right: zoom region with transit upper limit and TTV limit of 3 minutes on the transiting planets. I see that an additional Earth-size planet injected at a orbital period of $P\sim12$ days would still be likely to exist as it likely would not violate the transit and TTV limits for planets b and c. Habitable zones are shown like in Figure~\ref{fig:tauCeti}.}
    \label{fig:HD219134_P}
\end{figure*}

\begin{figure*}[ht]
    \centering
    \includegraphics[width=0.32\linewidth]{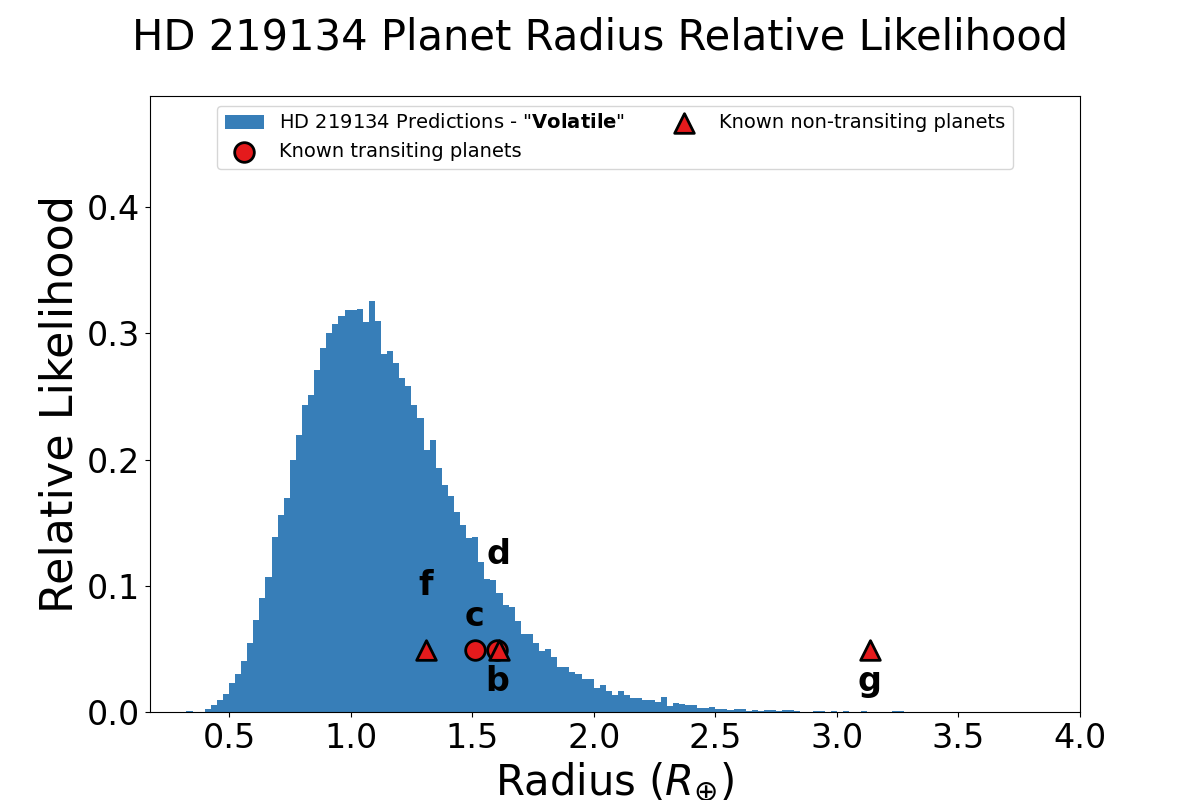}
    \includegraphics[width=0.32\linewidth]{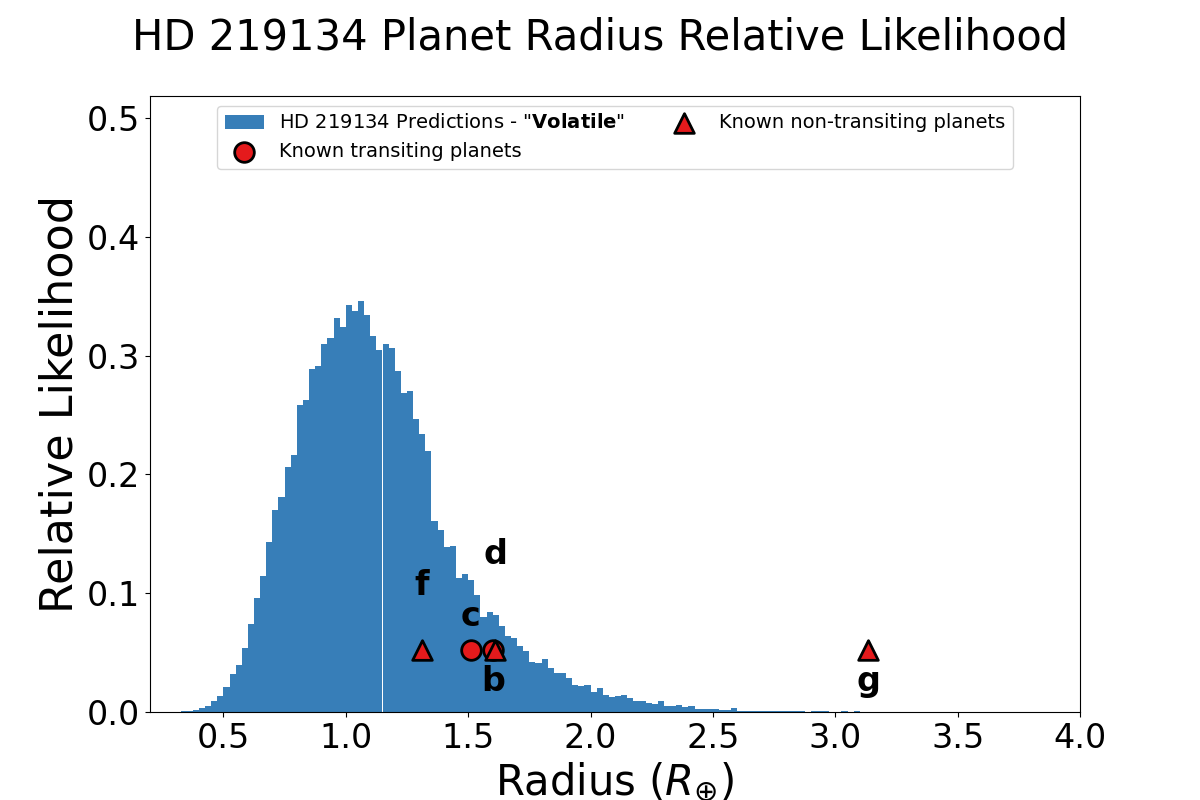}
    \includegraphics[width=0.32\linewidth]{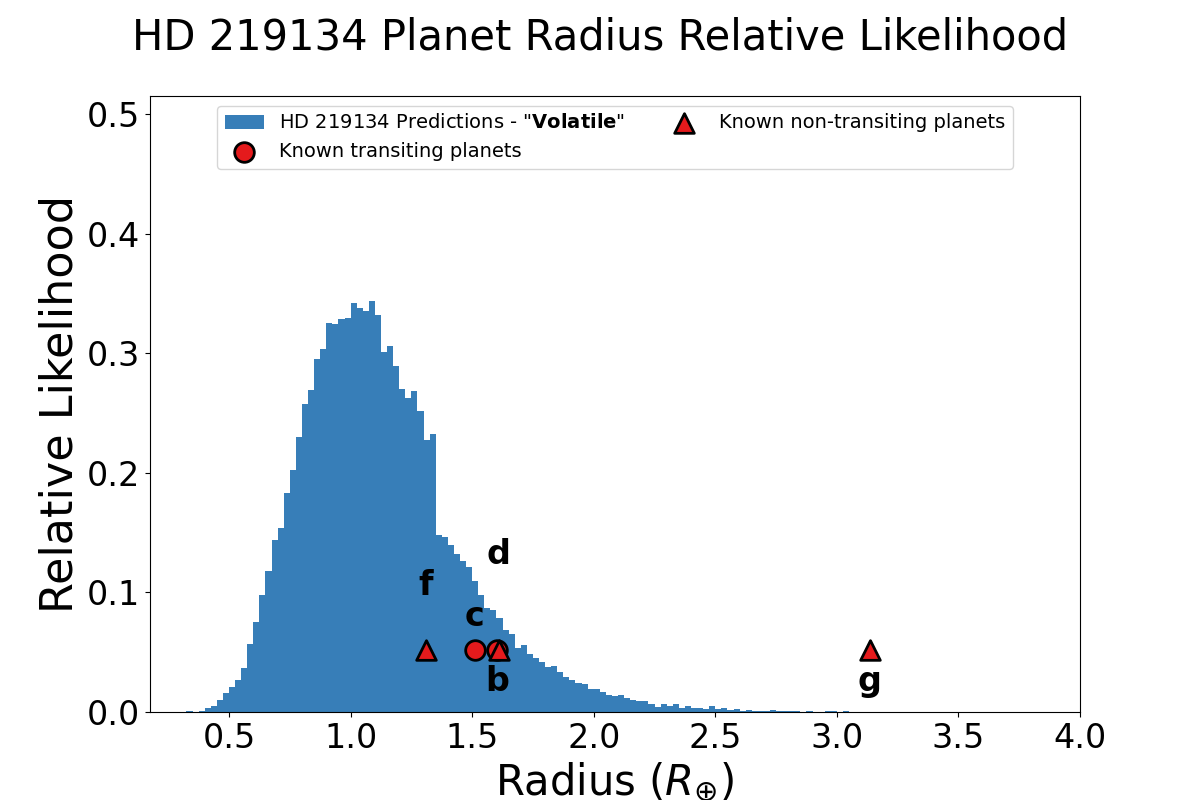}
    \caption{\tnt{} predictions in planet radius space for the HD 219134 system. Left: Full predictions for $0.1-4 R_{\oplus}$. Middle: predictions with transit upper limit equal to smallest found planet. Right: Predictions with transit upper limit and TTV limit of 3 minutes on the transiting planets. I see a relatively sharp drop off in planet radius at the the transit limit when it is included, as only some of the planets of those sizes would transit. However, the TTV limit does not noticeably change the predictions for planet radius.}
    \label{fig:HD219134_R}
\end{figure*}

Transiting planets b and c were also observed by TESS as TOI-1469.01 and .02. The transits themselves show no significant TTV signals in a variety of studies over nearly a decade of ephemeris modeling \citep[e.g.][]{Motalebi2015,Gillon2017a,Kokori2023}. This places a tight upper limit on the presence of additional planets as large or larger than the current transiting planets in the inner system, as well as the TTVs for the transiting planets, as TESS could have down to $\sim$3-minute TTV sensitivity for the targeted brighter host stars with its two-minute cadence \citep[e.g.,][]{Hadden2019}. Here I test how the likelihood of additional planets in the HD 219134 system changes with transit limits of $1.33 R_{\oplus}$ and TTV limits of 3 minutes for the inner two planets, and show the results in Figure~\ref{fig:HD219134_P} and Figure~\ref{fig:HD219134_R}. I note that this analysis is done now with consideration of the gas giant planet with an orbital period of $P\sim5-6$ years, unlike the analysis from \citetalias{Dietrich2021}. This planet falls outside the range of the robust Kepler statistics in both orbital period and planet size, but it may influence the stability of the inner system and ability to host planets out to an orbital period of $\sim2$ years, so it is included here.

The presence of the transit and TTV limits does not noticeably affect the orbital period predictions from \tnt{}, which still show a gap in between planets c and d that is likely to host a planet. The presence and most likely parameters of said additional planet would not induce noticeable TTVs on its transiting potential neighbor planet c, so the likelihood remains high of finding another planet there. However, the likelihood of finding a larger planet there drops noticeably, as if it were transiting it would likely have been found by now. Still, with planet f most likely not transiting, it is possible a planet larger than planet c could exist in the gap if it is not transiting.

In addition, there is a peak in the likelihood out beyond the outermost planet, which enters the habitable zone of HD 219134. This peak matches the signal seen at $\sim192$ days in the CPS/APF and the CLS data, but with conflicting determinations of the veracity of this signal I cannot confirm the presence of the planet there (especially since this prediction uses the 94-day period for planet g from \citet[][]{Vogt2015} and \citet[][]{Rosenthal2021}, and thus predicts an additional planet where \citet[][]{Hirsch2021} claim planet g is located). Both \citet[][]{Hirsch2021} and \citet[][]{Rosenthal2021} find a signal at 364 days but dismiss it as an annual systematic. Thus, I cannot rule out the peak in the likelihood at $\sim$185 days or the corresponding further peak at $\sim$360 days that occurs if I insert an additional planet at 185 days and run \tnt{} again (as in \citetalias{Dietrich2022}).

In summary, for HD 219134 I find that the predicted high likelihood in orbital period space of a planet existing between transiting planet c and likely non-transiting planet f is not significantly affected by the transit and TTV limit. This means that a planet is just as likely to be there in orbital period space, but it could possibly be smaller than expected, as shown by the decrease in larger planets in the expected planet radius distribution when the transit limit is included. The peak in likelihood beyond planet g is not significantly affected by the limits either, and it in the habitable zone of the star but is close to one half the orbital period of Earth, which has been believed to be the systematic cause for a signal near that period in the periodogram of the star.

\subsection{HD 20794} \label{subsec:res_prev_eEridani}

HD 20794, also known as 82 G. Eridani and e Eridani (and not to be confused with the third-closest naked eye star $\epsilon$ Eridani, which also has a known planet orbiting it), is a mid- to late-G dwarf located about 6 parsecs away from the Sun, making it the fifth-closest Sun-like star to the Solar System \citep[e.g.,][]{Pepe2011, Basant2022a}. This makes it potentially an interesting system from an astrobiological perspective to study for potential to host habitable planets, as it is also an old and relatively metal-poor star \citep[e.g.,][]{Valenti2005, PortodeMello2006, SuarezAndres2016, Shapiro2023}. HD 20794 also has an observed debris disk with a modeled semi-major axis of 24 AU and an inclination of $50^\circ$ \citep[][]{Kennedy2015}. I do not include the disk inclination value as a prior for the planet inclinations, both because there are no uncertainties on the value and also to show the effect of the RV limits on a system with no planetary inclination prior beyond ``non-transiting".

HD 20794 is known to host at least two known planets, although the number of planets and candidates in the system has changed across a decade of observations. The first study to ascertain the presence of planets was \citet[][]{Pepe2011}, which found evidence for three planet candidates: b (orbital period $P\sim18.3$ days), c ($P\sim40.1$ days), and d ($P\sim89.6$ days). An additional study in 2017 by \citet[][]{Feng2017a} found only weak evidence for planet c (which also may be tied to the stellar rotation period), but also provided parameters for three more planet candidates: e ($P\sim147$ days), f ($P\sim331$ days), and g ($P\sim11.9$ days). Further studies in 2023 used the now decade-plus of RV baselines and only found convincing evidence for planets b and d, but neither confirming nor refuting planet candidates e, f, and g \citep[][]{Cretignier2023, Laliotis2023}. However, an additional candidate exterior to all the known planets and candidates was found by \citet[][]{Cretignier2023}, with an orbital period of $P\sim645$ days and thus located beyond the outer reaches of the habitable zone, although with a moderate eccentricity of $\sim0.4$ it is partially located in the habitable zone near periastron.

In a previous \tnt{} analysis by \citet[][]{Basant2022a}, the different sets of planets and candidates found by \citet[][]{Pepe2011} and \citet[][]{Feng2017a} were treated in an iterative way, with anywhere from 3 to all 6 planets and candidates treated as true for determining where to likely find another planet. When only planets b, d, and e were considered true, the peaks in the likelihood corresponded to the orbital periods that matched planets c, f, and g, which were then added iteratively to the system with subsequent predictions matching the original. In addition, when planet f was considered true, the system was dynamically packed out to f and the likelihood for an additional planet peaked at 611 days, with the 68\% likelihood range from 548 to 733 days. Thus, the new planet candidate found in RV by \citet[][]{Cretignier2023} matches the predictions from the hypotheses where planet f was considered true, and the planetary system in general would likely be fully dynamically packed and relatively uniform in period ratios out to the outer reaches of the star's habitable zone.

\begin{figure*}[ht]
    \centering
    \includegraphics[width=0.32\linewidth]{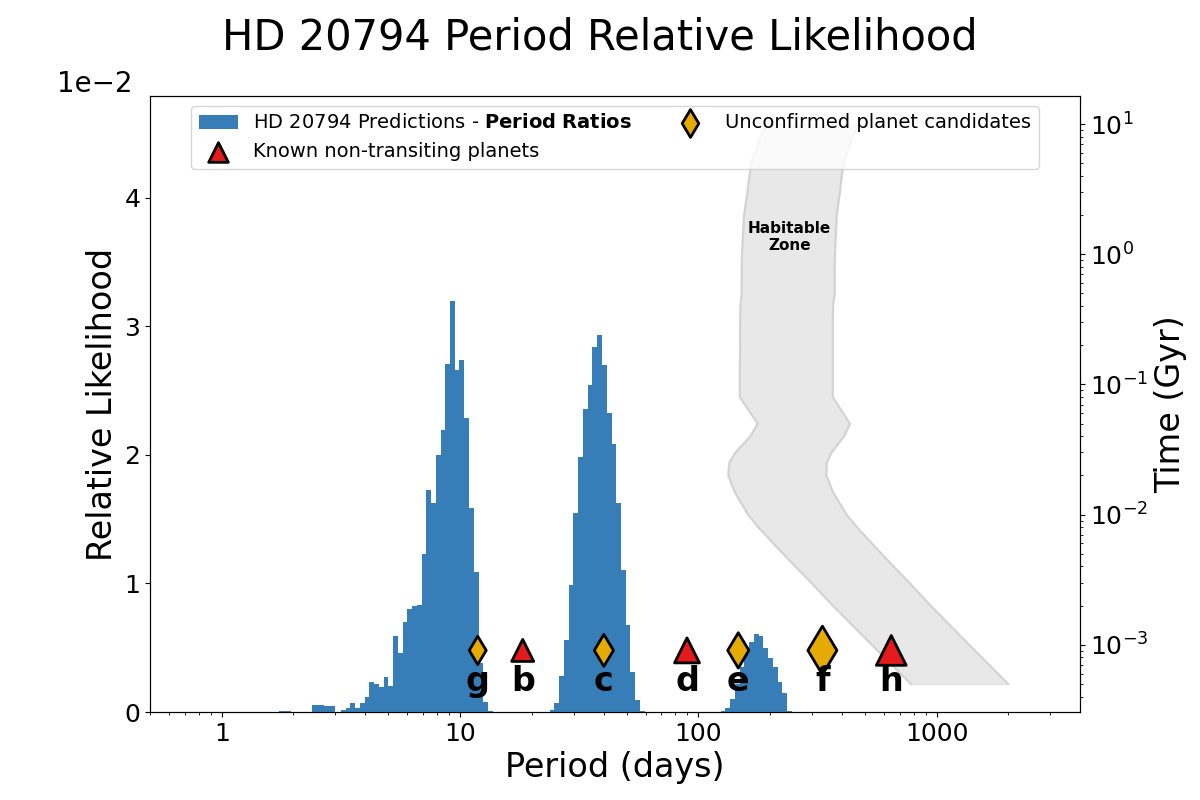}
    \includegraphics[width=0.32\linewidth]{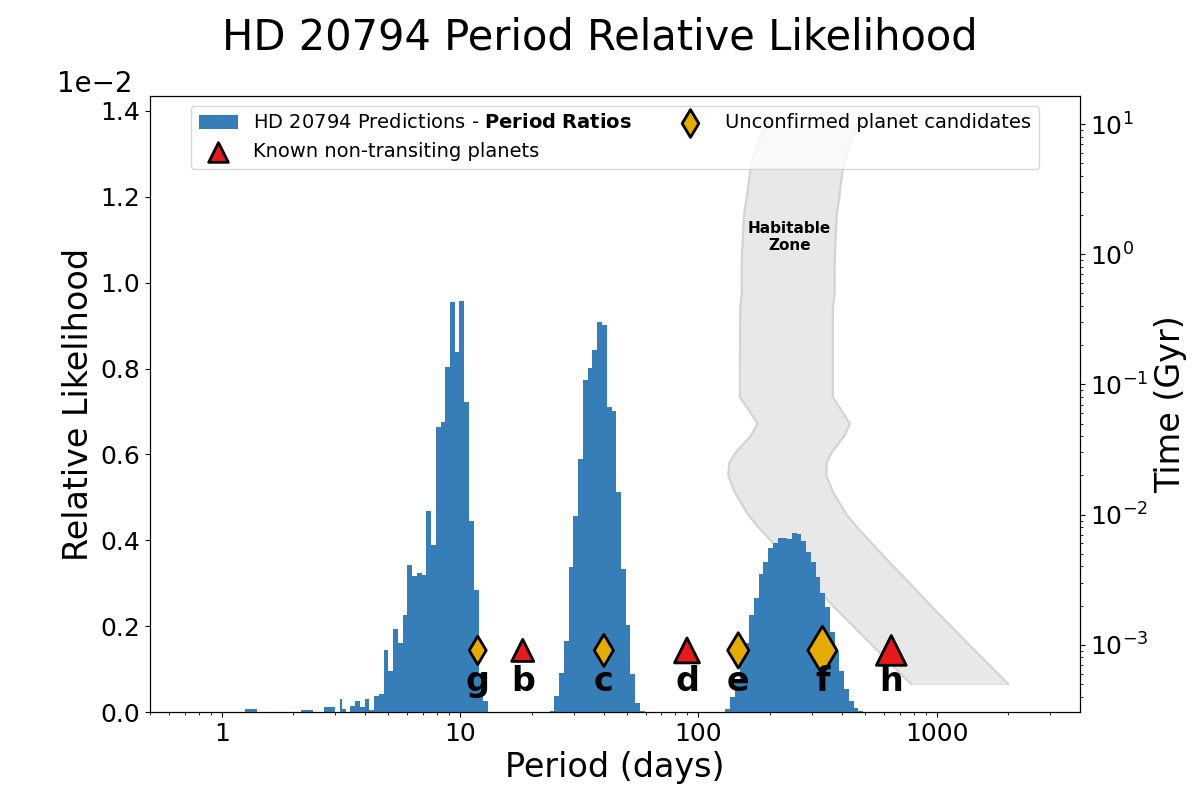}
    \includegraphics[width=0.32\linewidth]{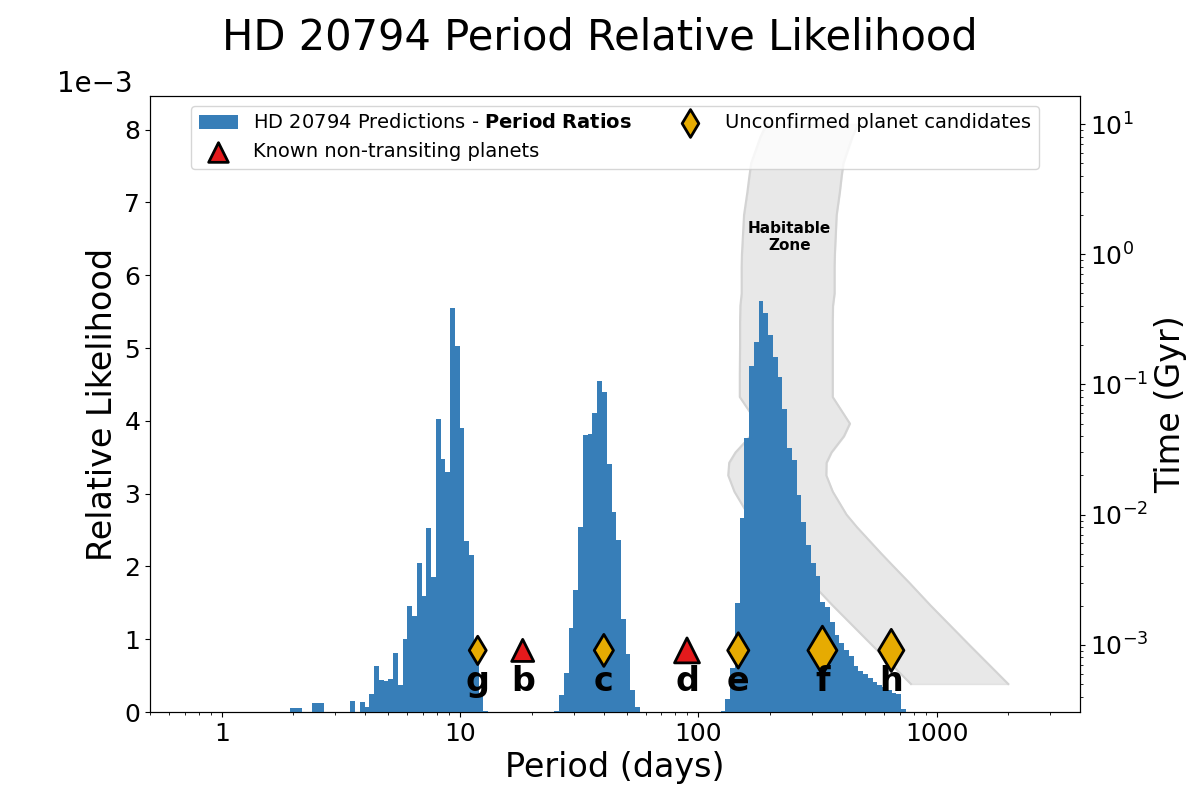}\\
    \caption{\tnt{} predictions in orbital period space for HD 20794 including planet h with an eccentric orbit (left), a circular orbit (center), and excluding it (right). Planets b and d are shown as confirmed as they are found in all four analyses, whereas planet candidates c \citep[][]{Pepe2011} and e, f, and g \citep[][]{Feng2017a} are shown as unconfirmed, where \tnt{} does not know of their possible existence or parameters. Habitable zones are shown like in Figure~\ref{fig:tauCeti}.}
    \label{fig:HD20794}
\end{figure*}

\begin{figure*}[ht]
    \centering
    \includegraphics[width=0.49\linewidth]{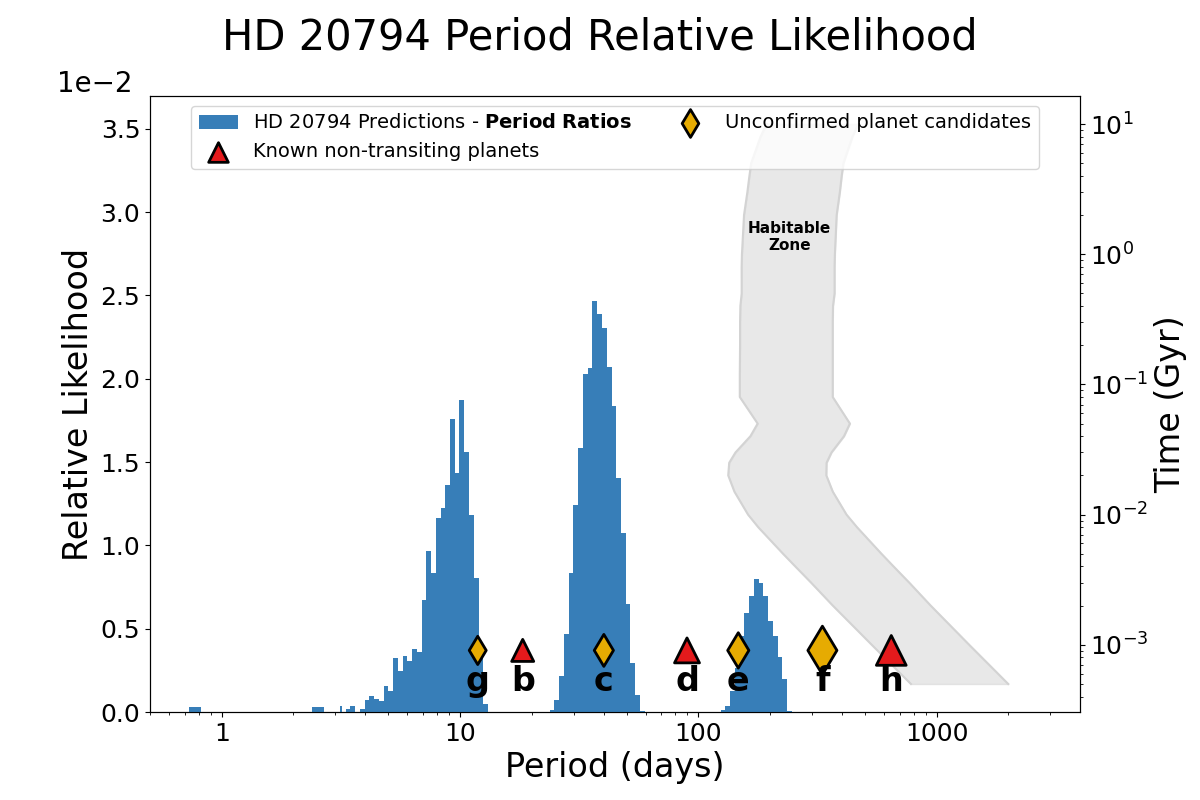}
    \includegraphics[width=0.49\linewidth]{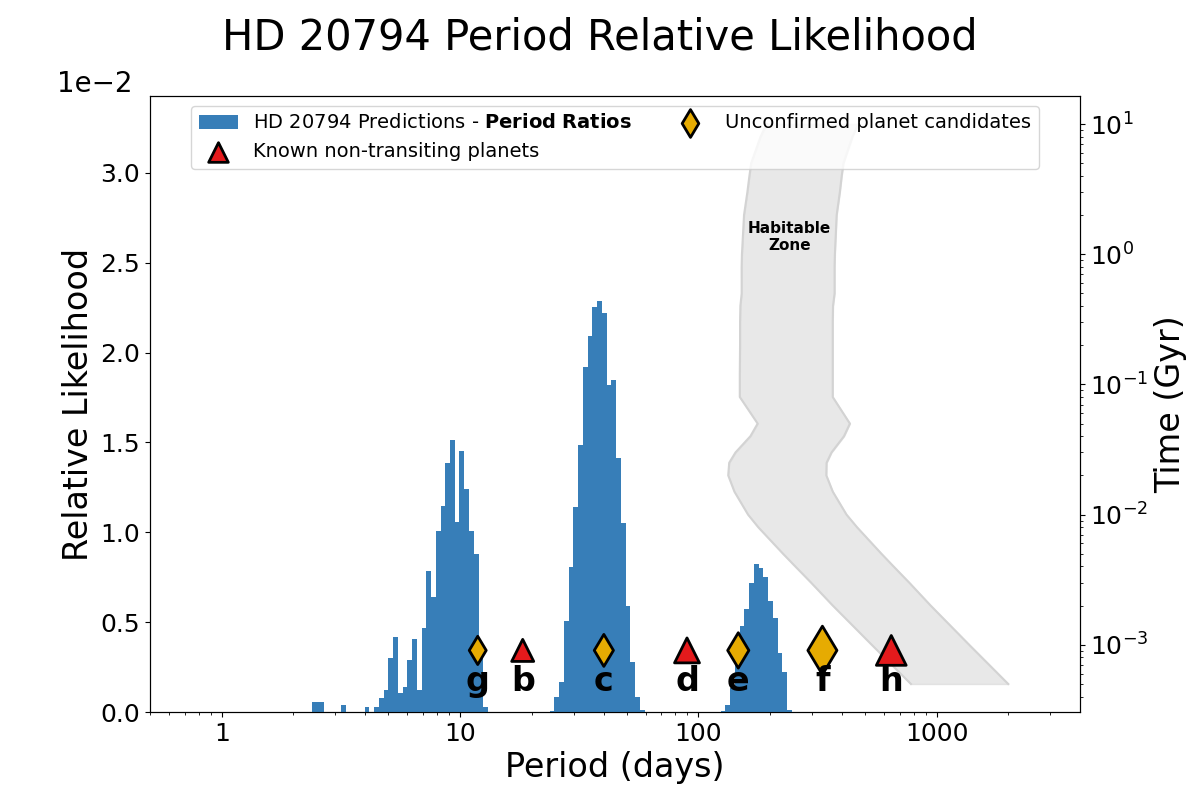}\\
    \includegraphics[width=0.49\linewidth]{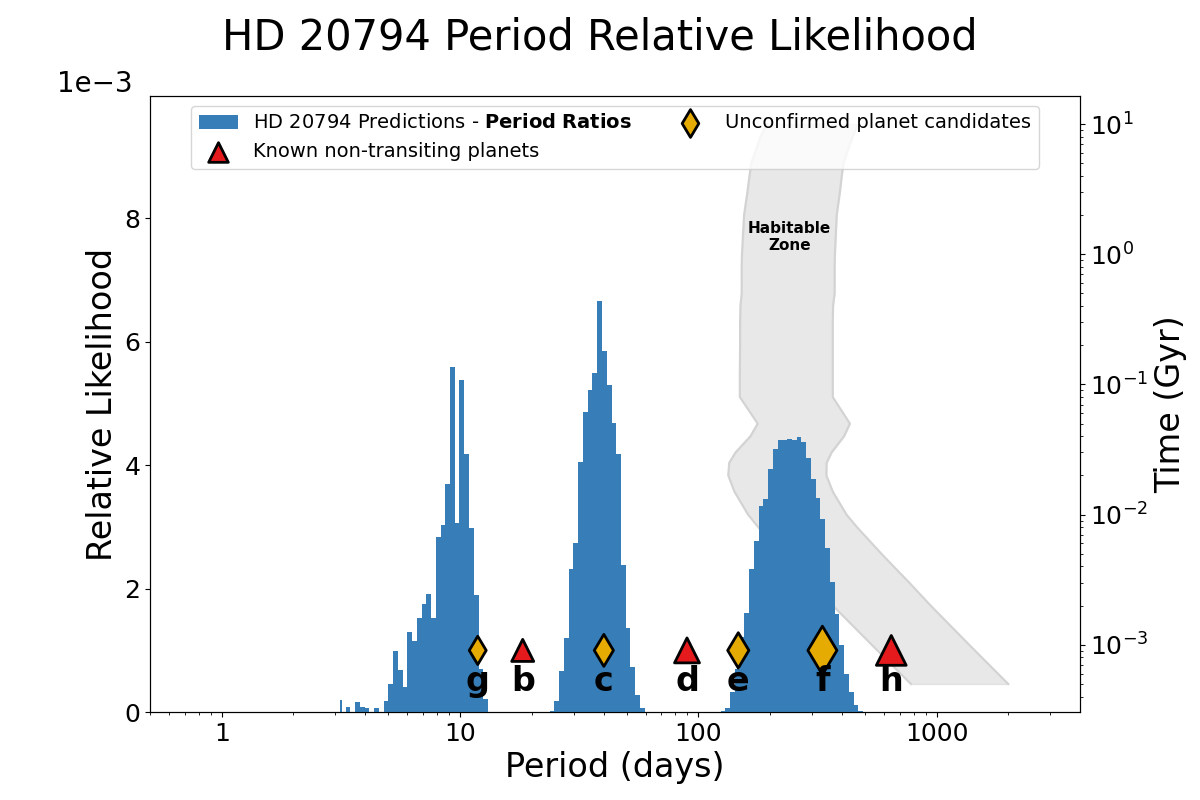}
    \includegraphics[width=0.49\linewidth]{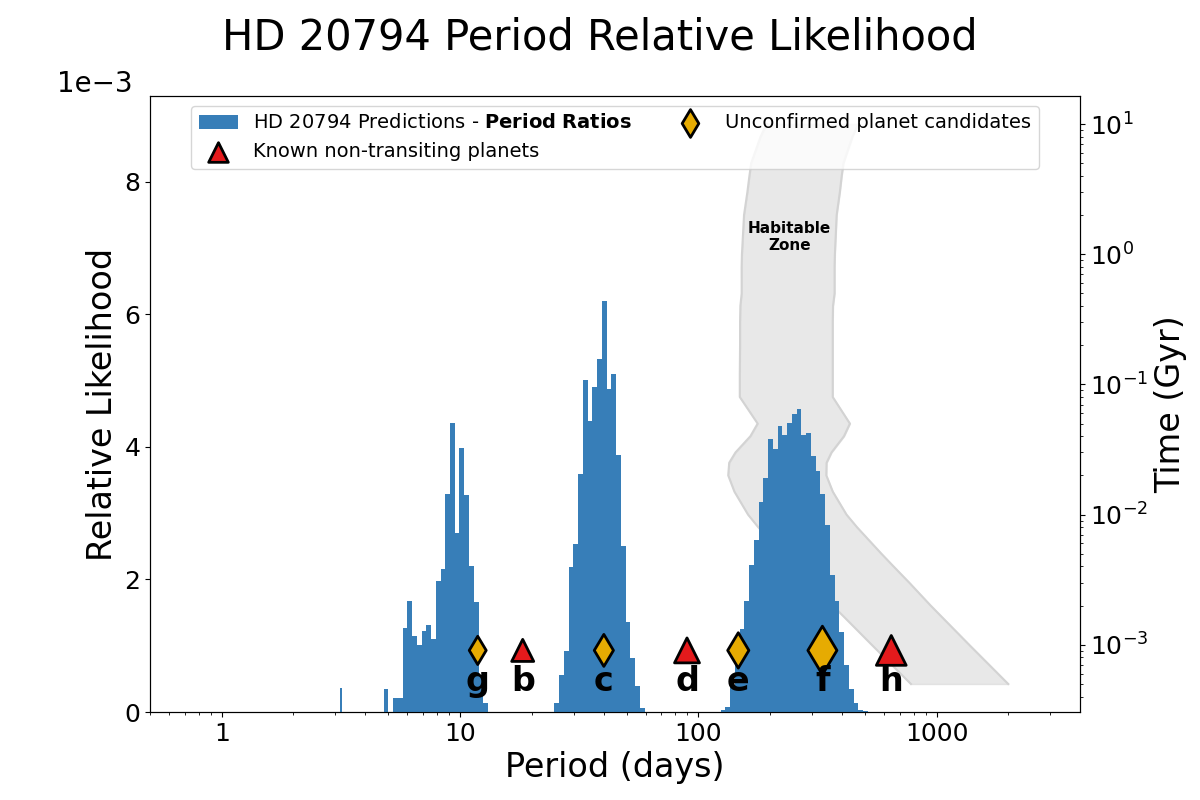}\\
    \includegraphics[width=0.49\linewidth]{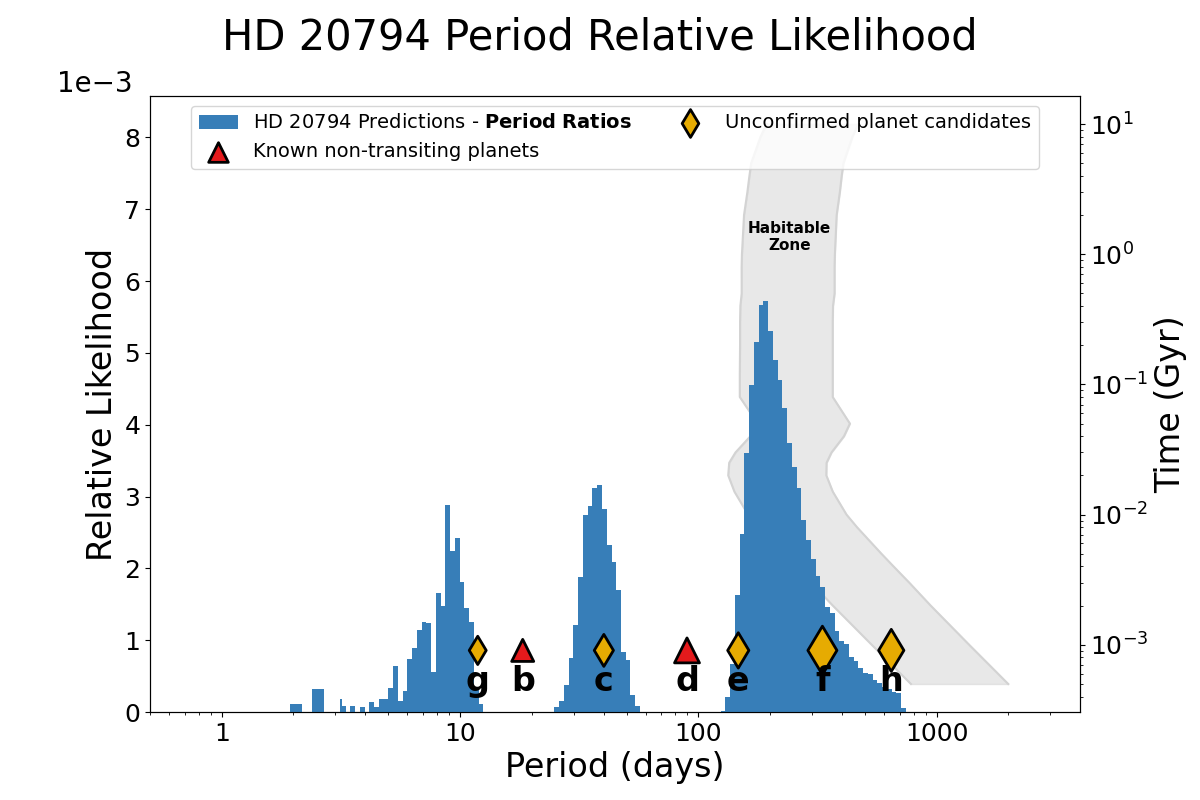}
    \includegraphics[width=0.49\linewidth]{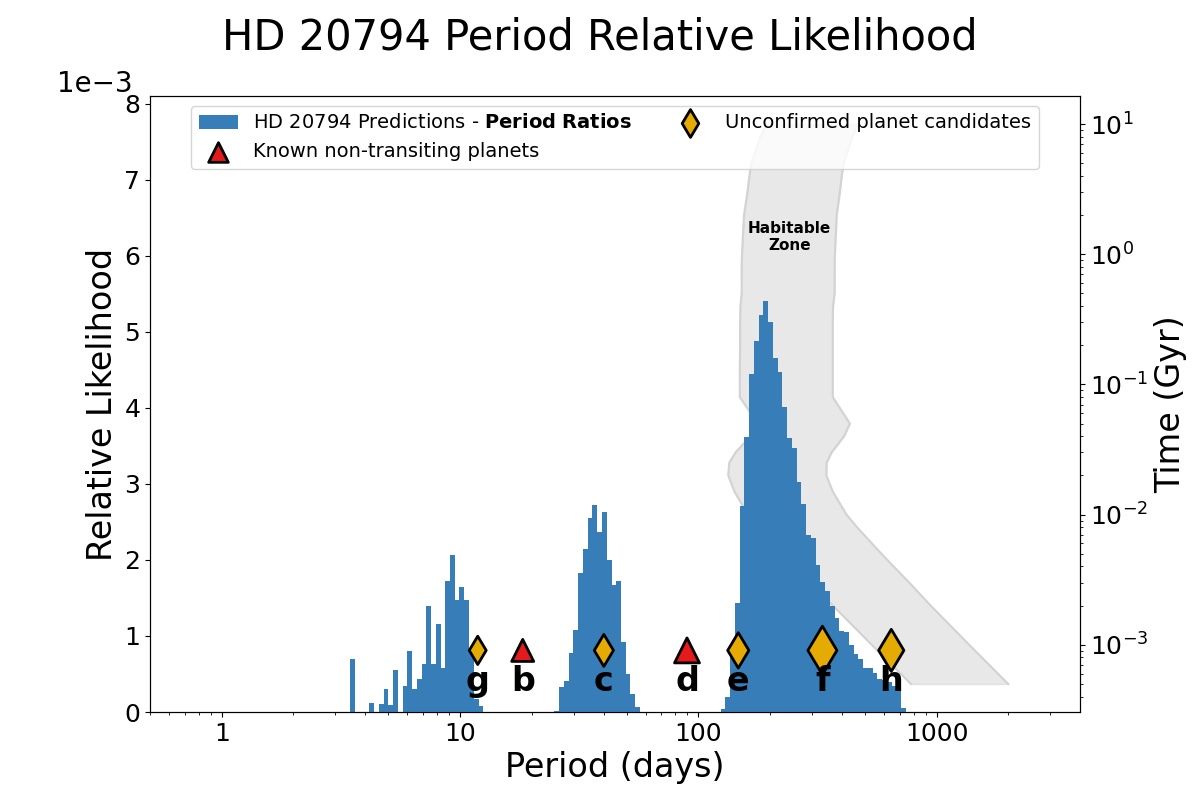}\\
    \caption{\tnt{} predictions in orbital period space for HD 20794 where the RV limit is 60 cm/s (left) and 25 cm/s (right) for the system with the 644-day planet from \citet[][]{Cretignier2023} on an eccentric orbit (top), a circular orbit (middle), and without it (bottom). Both the 60 cm/s limits and 25 cm/s limits do not significantly change the predicted values from those made without limits (see Figure~\ref{fig:HD20794}) except for slightly reducing the likelihood near the innermost planet candidate g. Habitable zones are shown like in Figure~\ref{fig:tauCeti}.}
    \label{fig:HD20794rvlim}
\end{figure*}

Here I use the two-planet system with the most consistent values from \citet[][]{Pepe2010, Feng2017a, Cretignier2023} as the base set, but including the new candidate from \citet[][]{Cretignier2023} in an alternate hypothesis (see Figure~\ref{fig:HD20794}). I then test the RV limits placed on this system of 60 cm/s (obtained from the lowest confirmed RV semi-amplitude from all of \citet[][]{Pepe2011}, \citet[][]{Feng2017a}, \citet[][]{Laliotis2023}, and \citet[][]{Cretignier2023}), as well as the residuals limit placed by \citet[][]{Cretignier2023} of 25 cm/s, to see if the latter limit disfavors any of the different candidates from \citet[][]{Feng2017a} for which no conclusive evidence in favor has yet been found. I show the results of these analyses in Figure~\ref{fig:HD20794rvlim}.

I find that the 60 cm/s limit lowers the likelihood of finding an additional planet between 10-100 days (including the orbital periods of planet candidates g and c) only in the case of h not being real, as the relative likelihood between the right panel of Figure~\ref{fig:HD20794} and the bottom left panel of Figure~\ref{fig:HD20794rvlim} only changes significantly interior to 100 days. The 25 cm/s limit imposes the same limit on the likelihoods exterior to planet b as the 60 cm/s limit does, and additionally it further lowers the likelihood of planet candidate g near 10 days in all three hypotheses (planet h with a circular orbit, planet h with an eccentric orbit, and without planet h). However, it does not significantly further affect the likelihoods where planet candidate c would exist as well as anything exterior to planet d.

\begin{figure*}[ht]
    \centering
    \includegraphics[width=0.49\linewidth]{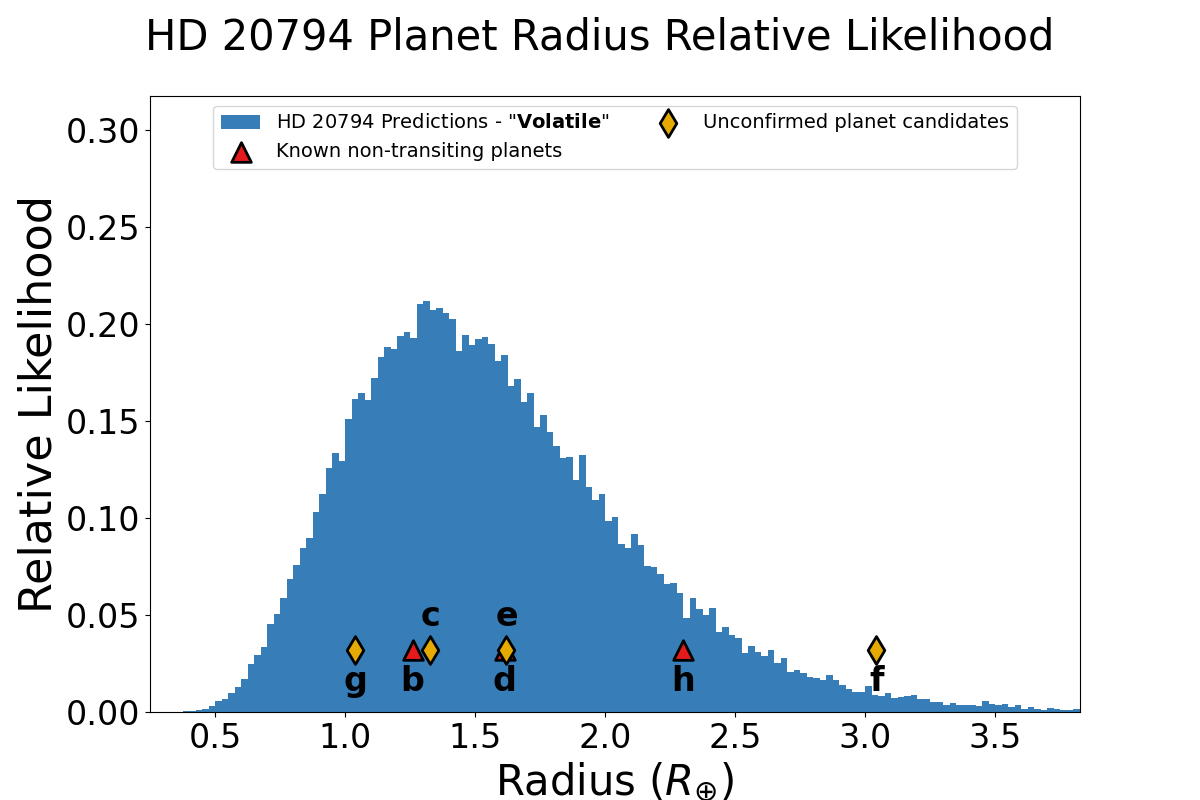}
    \includegraphics[width=0.49\linewidth]{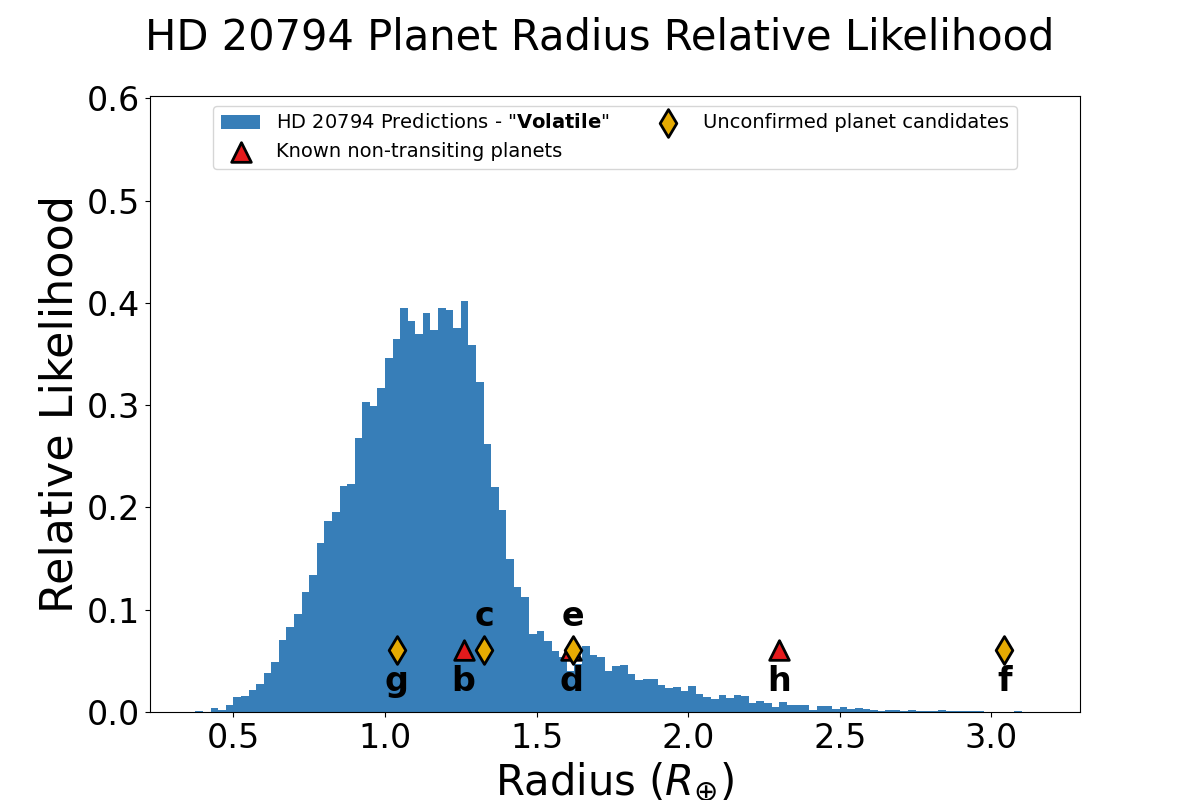}\\
    \includegraphics[width=0.49\linewidth]{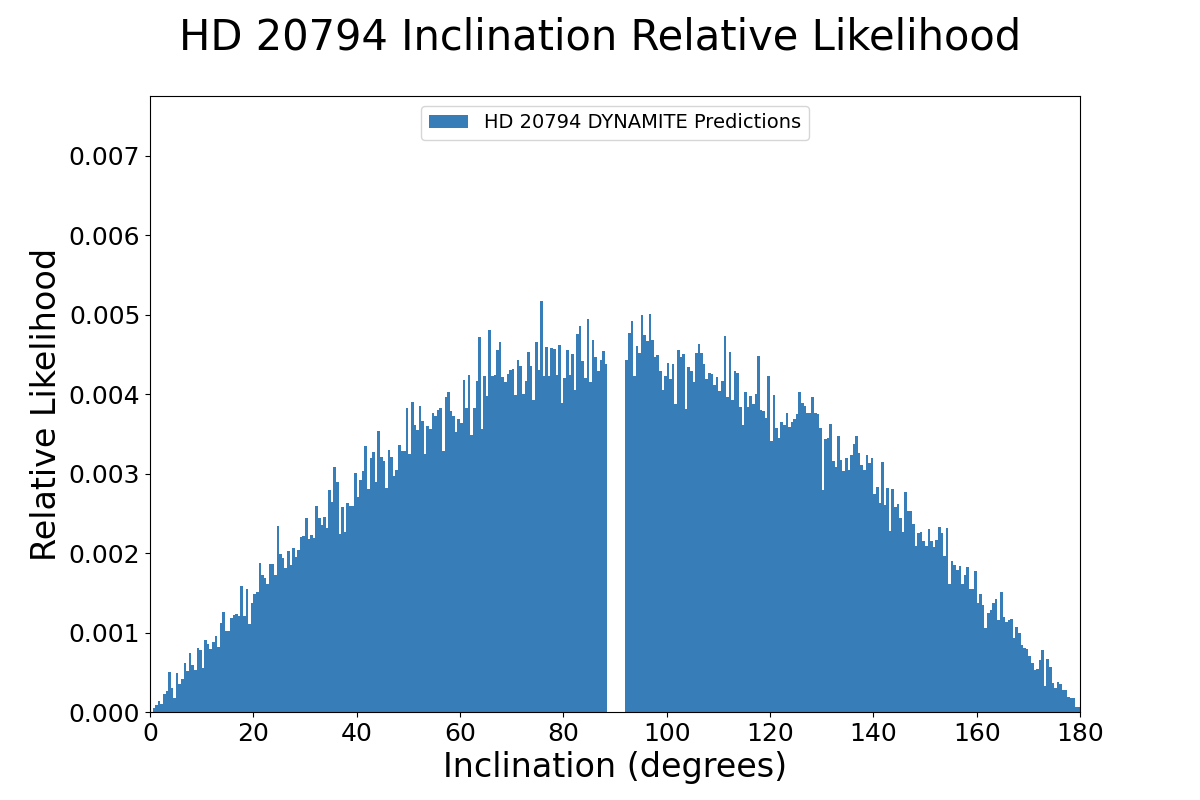}
    \includegraphics[width=0.49\linewidth]{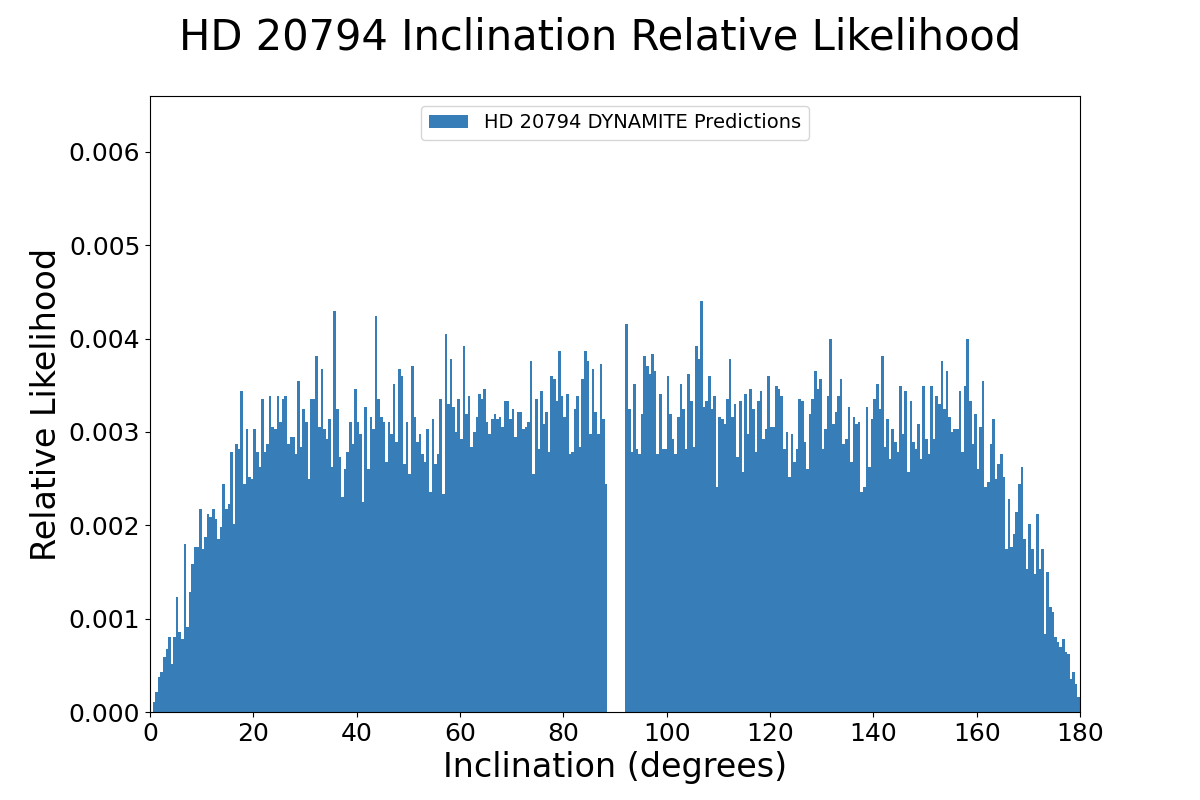}\\
    \caption{\tnt{} predictions for the planet radius (top) and orbital inclination (bottom) for no RV limits (left) and 25 cm/s RV limits (right). The planet radius distribution is calculated using the \citet[][]{Otegi2020} M-R relationship and a draw from the inclination distribution to get the hypothetical mass from the $m \sin i$ value. The strict RV limit significantly lowers the likelihood of higher-mass planets as well as planets closer to an edge-on orbital inclination, as both of these would have higher RV semi-amplitudes.}
    \label{fig:HD20794sizeinc}
\end{figure*}

The lower RV limits for HD 20794 also have a stronger effect on the predicted planet's size and orbital inclination, as shown in Figure~\ref{fig:HD20794sizeinc}. With the 25 cm/s limit, the likelihood of having planetary sizes above that of planet d is strongly suppressed compared to no RV limits. I note that the actual value shown assumes random draws from the inclination distribution shown, and thus may not be true if the inclination is constrained further (e.g., with an observed disk inclination), but the relative likelihood values are still accurate. In addition, the inclination distribution is flattened out throughout most of the available range (i.e., non-transiting), as planets of the same mass with inclinations closer to 90 degrees would have a higher RV semi-amplitude.

In summary, for HD 20794 I find that the RV upper limits, especially the 25 cm/s limit from \citet[][]{Cretignier2021}, significantly affect the presence of additional planets within 100 days and disfavors the existence of planet candidate g around 10 days. The 25 cm/s limit also strongly disfavors planets larger than planet b and finds them to be less likely to be found near 90 degrees inclination than a uniform distribution in $\cos i$. The presence of planet candidates e and f are much less dependent on the RV limits, as the likelihood in orbital period space is instead more strongly affected by the orbital eccentricity of (and therefore the dynamical stability with) planet h.

\section{Discussion} \label{sec:discussion}

\subsection{Observational upper limits} \label{subsec:disc_ul}

In this study, I included observational upper limits that provide constraints on the presence of additional planets through the lack of signal, instead of the appearance of one. Of the limits tested in this study, RV upper limits were both the most prevalent and had a noticeable impact. The transit probability limits were mostly utilized in analyzing the HD 219134 system as it was the only system with already known transiting planets, and those helped constrain the inclination and size of the potential injected planet between the last known transiting planet and the first known non-transiting planet in the system. Similarly, the TTV limits were really only helpful for that specific region of high likelihood space between planets c and d in the HD 219134 system, as a potential additional planet there would likely only cause TTVs on planet c. Further work to test transit and TTV limits would best be done on Kepler or TESS multi-planet systems (see Section~\ref{subsec:disc_ns}).

In contrast, there were planets and planet candidates found via RV in all of these systems, and correspondingly there were RV upper limits measured for all of these systems. In addition, the longer baseline of decade-plus observations for these nearby quiet solar-type stars translates into a clean and low-scatter set of RV data points, which can provide even down to 25 cm/s precision on any remaining signals in the data \citep[e.g.,][]{Cretignier2023}. Many of the signals for the planets/candidates are also sub-m/s, highlighting the precision and sensitivity required to find and confirm these planet candidates, as additional data has often caused different planetary candidate signals to appear and disappear from significance \citep[e.g.,][]{Feng2017a, Feng2017b, Cretignier2021, Cretignier2023}. The biggest remaining question for RV studies is the removal of the Earth's annual period (and therefore observing seasonal) systematic signal. An unfiltered periodogram of radial velocity data spanning multiple years will often have a signal around one year from observing Earth's motion as it orbits the Sun as well as incomplete phase coverage caused by observing seasons \citep[see e.g.,][]{Baluev2008, Lovis2010, Cretignier2023}. Most signals at these periods tend to be false positives due to the systematic, and potentially true RV signals from exoplanets around the target star could be missed for this reason. Thus, there exist regions in period space around G-type stars specifically where potentially habitable planets might be missed since their orbital period would be close to one year, likely requiring direct imaging to find and observe.

Multi-planetary systems from Kepler tend to show very high period and radius uniformity and very low orbital gap complexity \citep[i.e., ``peas in a pod"][]{Weiss2018, Millholland2017, Gilbert2020}, so if an additional planet is found in a given system, the statistics from Kepler prefer one that would fill in that gap and reduce the complexity. For the HD 219134 system, this trend manifests strongly through the high likelihood of finding another planet in the gap around 12 days, as this would be a significant deviation from the remaining gap sizes between planets. For \ensuremath{\tau} Ceti and HD 20794, however, this trend doesn't appear as strongly, due to the seemingly larger gaps between the RV-discovered planets. The additional unconfirmed candidates found respectively by \citet[][]{Tuomi2013} and \citet[][]{Feng2017a} for these two systems, along with the predicted planet at a near-year period for \ensuremath{\tau} Ceti, would likely dynamically pack the systems and reduce gap complexity. Indeed, the planet candidate with a 644-day orbital period reported by \citet[][]{Cretignier2023} would extend a hypothetically dynamically packed system beyond the habitable zone, as well as matching the prediction for an additional planet by \citet[][]{Basant2022a} relatively well. However, it is possible that these systems are fundamentally different than the Kepler population, showing that further work is necessary to determine if a population of such planetary systems exists.  Being able to combine planet occurrences from multiple surveys and multiple types of observations would help determine if there is an intrinsic difference between planet populations discovered via different methods.

\subsection{New systems to analyze} \label{subsec:disc_ns}

I chose to test observational upper limits on systems already analyzed by \tnt{} previously \citepalias{Dietrich2021, Dietrich2022, Basant2022a} to test the new capabilities and provide comparisons with the earlier predictions. However, there are troves of additional systems to study that would make use of these upper limits as well, including opening up the parameter space for \tnt{} analyses to single-planet systems.

The original study introducing \tnt{} ran the initial tests on the suite of TESS multi-planet systems known at the time \citepalias{Dietrich2020}. The inclinations for the TESS candidates are not released directly on the Exoplanet Archive, so for \citetalias{Dietrich2020} we used \texttt{EXOFAST} \citep[][]{Eastman2013} impact parameter estimates from the ExoFOP-TESS archive\footnote{\url{https://exofop.ipac.caltech.edu/tess/view_toi.php}} or assumed an uninformed prior on the impact parameter to calculate the expected inclination of the planet. A follow-up study (Turtelboom et al., in prep) on these systems has found that $\sim$40\% of these known multi-planet systems had additional planet candidates found in the intervening years. Of these, $\sim30\%$ matched the period and $\sim60\%$ matched the radius of the predicted planets from \citetalias{Dietrich2020}, while running the newest version of \tnt{} on the same dataset from 2020 resulted in $\geq50\%$ of the new planets matching the predictions from at least one of the period models and $\geq70\%$ matching the prediction from the radius model. This result, while not necessarily statistically significant compared to the null hypothesis, nevertheless shows that \tnt{} can help narrow down the parameter space to find additional planets in the TESS systems. Future work incorporating transit and TTV limits from TESS, which provides well-defined transit detection and TTV limits for inner planets due to its survey nature, will improve the accuracy of these predictions even further.

Finally, adding upper limits greatly increases the scope of \tnt{} by allowing it to test single-planet systems. Currently, systems with only one planet do not receive strong constraints on the likelihood of an additional planet, as \tnt{} would find that the preferred planetary parameters would be to match the size, inclination, and eccentricity while placing the planet either as close as dynamically possible (if assuming clustered periods model) or at roughly half or twice the orbital period (if assuming the period ratios model). Adding in constraints derived from observational upper limits on the presence of additional planets will alter the orbital period and size likelihood distributions, providing a stronger prediction on what to expect for the parameters (and observable signature) of an additional planet. As more than half of the currently known planetary systems are single planet detections, this would more than double the available parameter space for \tnt{}.

\subsection{Future direct imaging observations and missions}

One of the three most important science themes of the Astro2020 decadal survey was "to identify and characterize Earth-like extrasolar planets, with the ultimate goal of obtaining imaging and spectroscopy of potentially habitable worlds," \citep[][]{Astro2020}. Thus, the concept of the Habitable Worlds Observatory (HWO) was born from an amalgmation of two previous direct imaging concepts: the Habitable Exoplanets Observatory \citep[HabEx;][]{Gaudi2020} and the Large UV/Optical/Infrared Surveyor \citep[LUVOIR][]{LUVOIR2019}. An important precursor science analysis for HWO (not scheduled for launch until the 2040s) is to identify the highest-priority target stars with planetary systems that contain potentially habitable exoplanets. An initial list was created by \citet[][]{Mamajek2024} for the NASA Exoplanet Exploration Program (ExEP), and a uniform and robust preliminary input catalog \citep[HPIC;][]{Tuchow2024} from which target and scientific priorities can be assessed has also been created, but so far no planet or system has been listed as a definite target. The \ensuremath{\tau} Ceti, HD 219134, and HD 20794 systems were all listed in the ExEP precursor list as being potentially high-priority targets.

Due to the annual systematic of Earth's orbit around the sun (from both the Earth's own Doppler motion and the seasonal observability of targets), there is a gap where ground-based or Earth-orbit-based Doppler RV spectroscopy is not sensitive to planets. Therefore, RV limits are not necessarily as robust in these regimes and planets might be hidden to both RV and transits (if not inclined properly), thus needing direct imaging to be found. Current telescopes and instruments like MagAO-X at Magellan \citep[][]{Males2018} and SCExAO at Subaru \citep[][]{Jovanovic2015} routinely deliver contrasts on the level of $10^{-5}$ at separations on the order of a few tens of milliarcseconds. However, resolving an Earth-like planet in the visible (near-IR) around one of these stars would need contrasts around $10^{-9}$ ($10^{-7}$) at the same separations. Thus, visible direct imaging with the next generation of extremely large telescopes, such as GMagAO-X at the GMT \citep[][]{Males2022} and MICADO and METIS at the E-ELT \citep[][]{Davies2021, Brandl2021} would not likely be able to detect an Earth-like planet around these nearby Sun-like stars except for in a handful of cases.

\section{Summary} \label{sec:summary}

In this study I have included observational upper limits from RV, transits, and TTVs to improve the accuracy and precision of the predicted planet parameters from planetary system analyses using the \tnt{} framework. The key points of the study are as follows:

\begin{enumerate}
    \item Observational upper limits provide more accurate constraints on the presence of additional predicted planets in a planetary system.

    \item For systems already analyzed by an earlier version of \tnt{} in previous studies in \citetalias{Dietrich2021} (\ensuremath{\tau} Ceti), \citetalias{Dietrich2022} (HD 219134), and \citetalias{Basant2022a} (HD 20794), the addition of RV limits removes some of the likelihood for the unconfirmed planet candidates with orbital periods near 10 days to be real, as their signals would already be seen within the applied RV limits.

    \item The likelihoods further out in orbital period space systems are still as significant as when predicted without upper limits, and some fall potentially in areas where the Earth's orbital period and the natural observing seasons could interfere with the ability for RV to fully probe those regions.

    \item Transit and TTV limits had a narrower impact and were not found to significantly change the period predictions for the transiting planets and predicted additional planet in the HD 219134 system, but did constrain the planet size distribution.

    \item Direct imaging with the next-generation ELTs for some targets, and then with future space telescopes, are the best bet to determine what types of planets exist in the habitable zone around nearby solar-type stars.
\end{enumerate}

Combining data on both the presence of planets as well as the lack thereof in planetary systems helps to understand the underlying occurrence rates, allowing for even more accurate models of planetary system architectures. Multiple methods of planet detection and characterization are also able to contribute to these statistics, so modeling that includes all available data will be the most accurate in reproducing expected planetary systems. This will provide an additional benefit in the search for potentially habitable worlds and extraterrestrial biosignatures.

\begin{acknowledgments}
I would like to thank D\'aniel Apai and Kevin Hardegree-Ullman for fruitful conversations on the analysis and discussions. The results reported herein benefited from collaborations and/or information exchange within the program “Alien Earths” (supported by the National Aeronautics and Space Administration under Agreement No. 80NSSC21K0593) for NASA’s Nexus for Exoplanet System Science (NExSS) research coordination network sponsored by NASA’s Science Mission Directorate. Exoplanet datasets are made available by the NASA Exoplanet Science Institute at IPAC, which is operated by the California Institute of Technology under contract with the National Aeronautics and Space Administration. I acknowledge use of the software packages NumPy \citep[][]{Harris2020}, SciPy \citep[][]{Virtanen2020}, Matplotlib \citep[][]{Hunter2007}, REBOUND \citep[][]{Rein2012, Rein2019}, and TTVFaster \citep[][]{Agol2016a, Agol2016b}. The citations in this paper have made use of NASA’s Astrophysics Data System Bibliographic Services.
\end{acknowledgments}


\bibliography{main}{}
\bibliographystyle{aasjournal}



\end{document}